\documentclass[
final,
onefignum,
onetabnum,
a4paper
]{siamart190516}

\usepackage[a4paper]{geometry}

\usepackage{blebbing}
\usepackage{amsfonts}
\usepackage{graphicx}
\usepackage{mathmacros}
\usepackage{blebbing}
\usepackage{colonequals}
\usepackage{gnuplot-lua-tikz}
\usepackage{subcaption}
\usepackage{booktabs}
\usepackage{tikz}
\usetikzlibrary{arrows}
\usetikzlibrary{calc}
\ifpdf
  \DeclareGraphicsExtensions{.eps,.pdf,.png,.jpg}
\else
  \DeclareGraphicsExtensions{.eps}
\fi

\newcommand{\couplingIntCortex}{C_{\interfTwo{}}}
\newcommand{\couplingIntMembr}{C_{\interfOne{}}}
\newcommand{\pfCouplingIntCortex}[1][]{#1{C}_{#1{\cortexPhaseFieldSym}}}
\newcommand{\pfCouplingIntMembr}[1][]{#1{C}_{#1{\membrPhaseFieldSym}}}
\newcommand{\totalEnergyFunct}[2][]{\ONEARGOP{#1{U}}{#2}}
\newcommand{\boltzmConst}{k_B}
\newcommand{\temperature}{T}
\newcommand{\charEnergyLen}{\delta}
\newcommand{\koff}{k_{\text{off}}}
\newcommand{\totalSpecDens}{\rho_0}

\newcommand{\fdVelocitySym}[1]{\hat{\velocitySym}^{#1}}

\newcommand{\fdMembrPhaseFieldSym}[1]{\hat{\membrPhaseFieldSym}^{#1}}

\newcommand{\fdGenGlChemPotSym}[1]{\hat{\genGlChemPotSym}^{#1}}

\newcommand{\fdGenWillmChemPotSym}[1]{\hat{\genWillmChemPotSym}^{#1}}

\newcommand{\fdSpeciesOneSym}[1]{\hat{\pfSpeciesOne{}{}}^{\negthinspace#1}}

\newcommand{\fdSpeciesTwoSym}[1]{\hat{\pfSpeciesTwo{}{}}^{\negthinspace#1}}

\newcommand{\fdPressureSym}[1]{\hat{\pressureSym}^{#1}}

\newcommand{\discrGlChemPot}[3][]{%
	\ifthenelse{\isempty{#2}}{%
      	\ifthenelse{\isempty{#3}}{%
		  	\glGenChemPot[\schemeSubpscr{\meshSize}{#1}]{}{}%
        }{%
		  	\glGenChemPot[\schemeSubpscr{\meshSize}{#1}]{\cdot,#3}{}%
        }%
    }{%
      	\ifthenelse{\isempty{#3}}{%
		  	\glGenChemPot[\schemeSubpscr{\meshSize}{#1}]{#2,\cdot}{}%
        }{%
		  	\glGenChemPot[\schemeSubpscr{\meshSize}{#1}]{#2,\cdot}{}%
        }%
    }%
}
\newcommand{\discrGlGenChemPot}[3][]{%
  	\ifthenelse{\isempty{#2}}{%
      	\ifthenelse{\isempty{#3}}{%
		  	\glGenChemPot[\schemeSupscr{#1}]{}{\pfSpontMeanCurv}%
        }{%
		  	\glGenChemPot[\schemeSupscr{#1}]{\cdot,#3}{\pfSpontMeanCurv}%
        }%
    }{%
      	\ifthenelse{\isempty{#3}}{%
		  	\glGenChemPot[\schemeSupscr{#1}]{#2,\cdot}{\pfSpontMeanCurv}%
        }{%
		  	\glGenChemPot[\schemeSupscr{#1}]{#2,#3}{\pfSpontMeanCurv}%
        }%
    }%
}

\newcommand{\meshSize}{h}
\newcommand{\timeStepSz}{\tau}

\newcommand{\cahnNum}{\mathrm{Cn}}

\newcommand{\capillaryNum}{\mathop{\mathrm{Ca}}}
\newcommand{\transpNumBend}{\mathop{\mathrm{TBe}}}
\newcommand{\transpNumCpl}{\mathop{\mathrm{TCp}}}
\newcommand{\relNumBend}{\mathop{\mathrm{RBe}}}
\newcommand{\relNumCpl}{\mathop{\mathrm{RCp}}}
\newcommand{\pecletNum}{\mathop{\mathrm{Pe}}}

\newlength{\squareLen}
\newlength{\cortexRad}
\newlength{\membraneRad}
\newlength{\linkerFootRad}
\newlength{\legendMargin}
\newlength{\legendHeight}

\undef\indicatePhaseField

\newsiamremark{remark}{Remark}
\newsiamremark{hypothesis}{Hypothesis}
\crefname{hypothesis}{Hypothesis}{Hypotheses}
\newsiamthm{claim}{Claim}

\headers{}{}

\title{A Diffuse Interface Model for Cell Blebbing Including
Membrane-Cortex Coupling with Linker Dynamics\thanks{Submitted to the editors DATE.
\funding{This work was funded by the DFG through the RTG~2339.}}}

\author{Philipp Werner%
  	\thanks{Department of Mathematics, Friedrich-Alexander-Universität Erlangen-Nürnberg, 
    91058 Erlangen, Germany (\email{philipp.werner@fau.de}).}
\and Martin Burger%
  	\thanks{Department of Mathematics, Friedrich-Alexander-Universit\"at Erlangen-N\"urnberg, 
    91058 Erlangen, Germany (\email{martin.burger@fau.de}).}
\and Florian Frank%
  	\thanks{Department of Mathematics, Friedrich-Alexander-Universit\"at Erlangen-N\"urnberg, 
    91058 Erlangen, Germany (\email{florian.frank@fau.de}).}
\and Harald Garcke%
  	\thanks{Fakult\"at f\"ur Mathematik, Universit\"at Regensburg, 93040 Regensburg, Germany
    (\email{harald.garcke@mathematik.uni-regensburg.de}).}
}

\usepackage{amsopn}

\ifpdf
\hypersetup{
  pdftitle={A Diffuse Interface Model for Cell Blebbing Including
Membrane-Cortex Coupling with Linker Dynamics},
  pdfauthor={Werner, P. and Burger, M. and Frank, F. and Garcke, H.}
}
\fi

\begin{document}

\maketitle

\begin{abstract}
The aim of this paper is to develop suitable models for the phenomenon of cell blebbing, which allow for computational predictions of mechanical effects including the crucial interaction of the cell membrane and the actin cortex. For this sake we resort to a two phase-field model that uses diffuse descriptions of both the membrane and the cortex, which in particular allows for a suitable description of the interaction via linker protein densities. 

Besides the detailed modelling we discuss some energetic aspects of the models and present a numerical scheme, which allows to carry out several computational studies. In those we demonstrate that several effects found in experiments can be reproduced, in particular bleb formation by cortex rupture, which was not possible by previous models without the linker dynamics.
\end{abstract}

\begin{keywords}
 Cell blebbing, phase-field model, interacting interfaces, fluid-structure interaction.
\end{keywords}

\begin{AMS}
  68Q25, 68R10, 68U05
\end{AMS}

\section{Introduction}
    The formation of special membrane protrusions in eukaryotic cells has drawn the attention
    of biologists for a couple of decades now (cf. \cite{Charras2008, CharrasPaluch2008}),
    and is also referred to as ``cell blebbing''. 
    \begin{figure}
        \setlength{\squareLen}{6cm}
        \setlength{\cortexRad}{1cm}
        \setlength{\legendMargin}{3mm}
        \setlength{\legendHeight}{1cm}
        \pgfmathsetlength{\membraneRad}{\cortexRad+.25cm}
        \pgfmathsetlength{\linkerFootRad}{\cortexRad+1mm}
        \centering
        \begin{tikzpicture}[tips,scale=1]
            \draw[fill=white] (0,0) -- (\squareLen,0) -- 
                (\squareLen,\squareLen) -- (0,\squareLen) -- (0,0);

            \draw[fill=cellColor] (\squareLen/2,\squareLen/2) circle (\membraneRad);
            \pgfmathsetlengthmacro\ax{\squareLen/2 + cos(135)*(\membraneRad+0.1cm)}
            \pgfmathsetlengthmacro\ay{\squareLen/2 + sin(135)*(\membraneRad+0.1cm)}       

            \draw[dotted] (\squareLen/2,\squareLen/2) circle (\cortexRad);
            \pgfmathsetlengthmacro\ax{\squareLen/2 + cos(45)*(\cortexRad-0.1cm)}
            \pgfmathsetlengthmacro\ay{\squareLen/2 + sin(45)*(\cortexRad-0.1cm)}       
            \pgfmathsetlengthmacro\ax{\squareLen/2 + cos(135)*(\cortexRad)}
            \pgfmathsetlengthmacro\ay{\squareLen/2 + sin(135)*(\cortexRad)}       
            \pgfmathsetlengthmacro\bx{\squareLen/2 + cos(135)*(\membraneRad)}
            \pgfmathsetlengthmacro\by{\squareLen/2 + sin(135)*(\membraneRad)}
            \draw[serif cm-serif cm] (\ax,\ay) -- (\bx,\by);
            \node[anchor=south east] at ($(\ax-2mm,\ay)$) {\footnotesize $ 10-20 $\,nm};

            \foreach \phi in {0,30,90,150,190,240,290}
                \pgfmathsetlengthmacro\ax{\squareLen/2 + cos(\phi)*\linkerFootRad}
                \pgfmathsetlengthmacro\ay{\squareLen/2 + sin(\phi)*\linkerFootRad}
                \draw[fill=yellow!70!red] (\ax,\ay) circle (1mm);

            \foreach \phi in {20,50,110,180,210,310}
                \pgfmathsetlengthmacro\ax{\squareLen/2 + cos(\phi)*\cortexRad}
                \pgfmathsetlengthmacro\ay{\squareLen/2 + sin(\phi)*\cortexRad}
                \draw[fill=green!50!black,rotate around={\phi+90:(\ax,\ay)}] (\ax,\ay) ellipse (2mm and 0.5mm);

            \draw ($(1cm,\squareLen-0.5cm)$) node[anchor=west] {
                \footnotesize extracellular matrix
            };
            \node at ($(\squareLen-1cm,\squareLen-1cm)$) {$\domain$};

            \draw ($(\squareLen/2,\squareLen/2)$) node[anchor=south] {
                \footnotesize cytosol
            };

            \pgfmathsetlengthmacro\ax{\squareLen/2 + cos(300)*(\membraneRad)}
            \pgfmathsetlengthmacro\ay{\squareLen/2 + sin(300)*(\membraneRad)}       
            \draw[serif cm-serif cm] ($(\squareLen/2,\squareLen/2)$) -- node[anchor=east] 
                {\footnotesize $ 2\mu\SIMeter $} (\ax,\ay);

            \draw[fill=yellow!70!red] ($(\legendMargin,\legendHeight)$) circle (1mm);
            \node[anchor=west] (lp) at ($(\legendMargin+2mm,\legendHeight)$) {\footnotesize linker proteins};
            \draw[fill=green!50!black] ($(\legendMargin,\legendHeight-0.5cm)$) ellipse (2mm and 0.5mm);
            \node[anchor=west] (mm) at ($(\legendMargin+2mm,\legendHeight-0.5cm)$) 
                {\footnotesize myosin motors};
            \draw[solid] ($(lp) + (2cm,0)$) -- ($(lp) + (2.5cm,0)$);
            \node[anchor=west] at ($(lp)+(2.7cm,0)$) {\footnotesize membrane};
            \draw[dotted] ($(mm) + (2cm,0)$) -- ($(mm) + (2.5cm,0)$);
            \node[anchor=west] at ($(mm)+(2.7cm,0)$) {\footnotesize cortex};
        \end{tikzpicture}
        \caption{A simple cell scheme.}
        \label{fig:introduction:cell scheme}
    \end{figure}
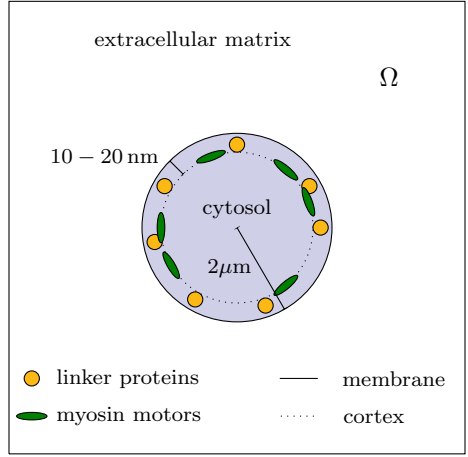            
    Cell blebbing has been observed during apoptosis \cite{Laster+1996},
    cytokinesis, and mitosis \cite{Boss1955, Nakatsuji+1986}, but also 
    as a means for locomotion \cite{Trinkhaus1980} in primordial germ cells or cancer cells.
    In order for a cell to move in a certain direction via blebbing, the site of bleb nucleation 
    towards that direction has to be specified.
    It has not yet been fully understood what triggers the formation of a bleb at 
    a specific cell site. However, theories about relevant influences exist and
    a mathematical model for studying the mechanisms that lead to bleb site 
    selection has been recently proposed in \cite{Dirks+2019}.
    In this work, we do not deal with this phase in the process of cell blebbing:
    our model targets the 
    process after bleb site selection and is aimed at investigating the mechanics of 
    bleb formation.
    Previous works on this topic are, e.g., \cite{Lim+2012, Strychalski+2013, 
    Strychalski+2015, Alert+2015, Alert+2016, Stinner+2020, Werner+2020}.
    Lim et al. \cite{Lim+2012} present a one-dimensional model focusing
    on bending and surface energy of the membrane and include another 
    important energy for the influence of so-called linker proteins (see below) that 
    also play a role in the development of membrane protrusions.
    The authors of \cite{Strychalski+2013, Strychalski+2015} deal with the
    interaction between membrane components and the surrounding fluid.
    In \cite{Alert+2015, Alert+2016} one-dimensional models which concentrate on the dynamics of the linker 
    protein density are investigated.
    Stinner et al. \cite{Stinner+2020} propose a finite element method (FEM) 
    for simulation of bleb heights on two-dimensional
    surfaces incorporating bending energy and linker effects as well.

    The aim of this paper is to propose a fully integrated three-dimensional model that takes
    forces of the surrounding fluid, linker protein effects as well as bending 
    and surface energies into account.
    In the following, we will treat bleb formation as a fluid-structure interaction problem with two 
    diffuse interfaces, the membrane and the cortex (see below), which are immersed in a fluid.
    On the cortex, we model the linker proteins with a density function and include their 
    forces on the membrane thus introducing a coupling between cortex and membrane.
    We will show how this coupling can be incorporated in a potential energy functional
    that also comprises surface and bending energy of the membrane and cortex.
    By leveraging Onsager's variational pinciple, we will arrive at a system of partial differential 
    equations (PDEs) with evolution laws for the diffuse interfaces as well as a Navier-Stokes 
    system for the fluid flow. Additionally, we will have a reaction-diffusion-type 
    PDE for the linker proteins concentrated on the cortex.

    \subsection{Notation} We denote the $n$-dimensional Lebesgue measure by $ \lebesgueM{n} $
    and the Hausdorff measure of Hausdorff dimension $m$ by $ \hausdorffM{m} $. The normal 
    of a surface $ \mathcal{S}$ is $ \normal[\mathcal{S}]{} $. The surface material 
    derivative is $ \matDeriv{}{}\rho $ and the surface gradient on a surface $ \mathcal{S}$ is
    denoted by $ \grad{\mathcal{S}}{}{} \rho $. The gradient for a Gateaux derivative of $ f $ in space 
    $ X $ is denoted $ \grad{}{X}{} f $. 
    $ \mathbb{P}_k $ denotes a simplicial element with polynomials of degree 
    {at most}~$k~$.
    Sobolev spaces of $ k$-times weakly differentiable, $ L^2 $ functions is 
    $ \sobolevHSet{k}{\domain}$. 
    The space of solenoidal functions is denoted $ \sobolevHSolSet[0]{1}{\domain} $.

\section{Modeling}
In the following we discuss the modelling of cell blebbing using diffuse interfaces. We start with a brief overview of the process before proceeding to a more mathematical level:

\subsection{Biological Background}
    Let us briefly clarify some basic parts of a cell and the principle processes 
    taking place when blebbing occurs.

    Fig.~\ref{fig:introduction:cell scheme} shows a simple cell scheme that 
    hints to the following components of a cell:
    \begin{itemize}
        \item The {\em cell membrane} (solid black circle)  is a bilayer of lipid molecules 
        and encloses the whole cell.
        \item The {\em cell cortex} (dotted black circle) is a network of         actin fibres, which are elastic and can be contracted by myosin motors 
        (green ellipses). 
        \item The {\em cytosol} is a fluidic material that fills up the whole cell
        and contains various different objects, which we do not model here explicitly, but within the fluid viscosity. 
        \item {\em Linker proteins} (orange small circles)    connect the cell membrane and 
        cell cortex.
        They are also elastic to a certain extent until they are overstretched and in 
        turn ripped apart. We model them as Hookean springs.
    \end{itemize}
    The cell is surrounded by a region called the extracellular matrix, which is also 
    filled with a fluid. In our model, we assume for simplicity a mixture of the cytosol 
    and the fluid in the extracellular matrix.

    The phenomenon of cell blebbing  may be outlined as follows (cf.\
    description in \cite{CharrasPaluch2008}):
    Initially, the cell cortex may be contracted by the myosin motors
    as a result of chemical reactions, or the cortex may be destroyed by external 
    influence.
    In turn, cytosol flows towards the cell 
    membrane (being either pushed by the cortex or pressed through a hole in case of 
    a destroyed cortex), which is then elongated forming a protrusion that 
    is called a bleb. This development is hindered by influences like surface tension and
    flexural rigidity of the cell membrane, as well as the linker proteins,
    which pin the cell membrane to the cell cortex.
    However, if the pressure resulting from the cell cortex contraction
    surpasses a critical value, the linker proteins break and the speed of bleb development
    changes qualitatively (cf.~\cite[p.~3]{Lim+2012}, \cite[p.~44]{Werner+2020}); this is called
    bleb nucleation.
    Subsequently, the phase of bleb retraction starts by reformation of the cortex.
    Furthermore, broken linker proteins are rebuilt and reconnect to the membrane. 
    In the end, the proteins pull the membrane towards the cortex and the bleb
    vanishes. This process is called `bleb retraction' or `bleb healing'.
\subsection{Phase Field Model}
    The cell with all its components is contained in a compact set $ \domain \subseteq \reals^3 $. 
    Our model comprises two diffuse layers, the cell membrane and cortex, both with width of order
    $ \phaseFieldParam > 0 $. They are represented by
    the phase field functions
    $ \funSig{\membrPhaseField{}{} ,  \cortexPhaseField{}{}}{
    \domain}{\reals} $
    that are supposed to approach the value $ -1 $ on the regions enclosed
    by the cell membrane and cell cortex, respectively,
    and $ 1 $ on the corresponding outer regions.
    In between these regions, the functions
    $ \membrPhaseField{}{} $, $ \cortexPhaseField{}{} $ interpolate smoothly
    and cross zero where the diffuse layers are to be expected.
    Phase field methods have been successfully applied in materials science \cite{Steinbach2009} and
    for two-phase fluid flow \cite{MR3916798}, where they have been motivated by thermodynamic theories of 
    phase separation. 

    The membranes surrounding
    eukaryotic cells are typically bilayers of lipid molecules. Assuming the fluid to be 
    lipophobic, say we are dealing with water, we expect the cell membrane to 
    show phase separation behaviour like, e.g., oil
    in water. For simplicity, we assume the same behaviour for the cell cortex, although
    the interaction of actin with water is more complex.

    \paragraph{The Canham--Helfrich Energy}
        Canham \cite{Canham1970} and Helfrich \cite{Helfrich1973} both 
        suggested a model for the energy associated to biomembranes.
        They basically came up with the following expression for a membrane
        $ \mathcal{M} $ as two-dimensional surface:
        \begin{equation*}
            \integral{\mathcal{M}}{}{
                \frac{\bendRig}{2}
                \left(\meanCurv{}{x}-\spontMeanCurv\right)^2
                +
                \elComprMod
            }{\hausdorffM{2}(x)}
            +
            \gaussBendRig
            \integral{\mathcal{M}}{}{
                \gaussCurv{}{x}
            }{\hausdorffM{2}(x)},
        \end{equation*}
        where $ \bendRig $ is the flexural rigidity, $ \spontMeanCurv $ is the spontaneous
        mean curvature (the mean curvature the membrane tends to in unconstrained situations),
        $ \elComprMod $ is the surface tension, and $\gaussBendRig $ is the Gaussian
        flexural rigidity of the membrane.
        The spontaneous curvature may be employed to reflect asymmetries in the membrane, 
        e.g., due to chemical influence (cf.~\cite[Sec.~2.5.2]{Seifert1997}).
        The Gaussian curvature is constant across a set of surfaces with the same 
        topologial genus. Hence, since we do not consider topological changes, 
        we may as well neglect the Gaussian bending term when varying 
        the energy functional.

        In order to employ the Canham--Helfrich energy in our phase field model,
        we need diffuse counterparts of the surface 
        $
            \integral{\mathcal{M}}{}{
                \elComprMod
            }{\hausdorffM{2}}
        $ 
        and bending
        $
            \frac{\bendRig}{2}
            \integral{\mathcal{M}}{}{
                \left( \meanCurv{}{x} - \spontMeanCurv \right)^2
            }{\hausdorffM{2}(x)}
        $ energies.
        It is well-known that the Ginzburg-Landau energy
        $$
            \ginzburgLandauEnergyFunct{\membrPhaseFieldSym}{}
            =
            \elComprMod
            \integral{\domain}{}{
                \frac{\phaseFieldParam}{2}\abs{\grad{}{}{}\membrPhaseFieldSym}^2
                +
                \frac{1}{\phaseFieldParam}
                \doubleWellPot{\membrPhaseFieldSym}
            }{\lebesgueM{3}}
            =
            \elComprMod
            \integral{\domain}{}{
                \ginzburgLandauEnergyDens{\membrPhaseFieldSym}{}
            }{\lebesgueM{3}}            
        $$
        $ \Gamma $-converges to the former as $\phaseFieldParam \downarrow 0$ (up to a constant factor, cf.~\cite{Modica1977,%
        Modica1987}).
        Moreover, it has been proven in \cite{Roeger+2006} (and independently
        in \cite{Nagase+2007} for two dimensions) that the functional 
        $$
            \pfGenWillmoreEnergyFunct{\membrPhaseFieldSym}{0}
            =
            \frac{1}{2\phaseFieldParam}\integral{\domain}{}{
                \left(
                    -\phaseFieldParam\laplacian{}{}{}\membrPhaseFieldSym + 
                    \frac{1}{\phaseFieldParam}
                    \doubleWellPot[\schemeSupscr{\prime}]{\membrPhaseFieldSym}
                \right)^2
            }{\lebesgueM{3}}
        $$
        $ \Gamma $-converges to the Willmore functional.
        In \cite{Du+2005}, this functional has been extended to the more 
        general case of spontaneous mean curvature such as
        $$
            \pfGenWillmoreEnergyFunct{\membrPhaseFieldSym}{\pfSpontMeanCurv}
            =
            \frac{1}{2\phaseFieldParam}
            \integral{\domain}{}{
                \left(
                    -\phaseFieldParam\laplacian{}{}{}\membrPhaseFieldSym 
                    + 
                    \left(
                        \frac{1}{\phaseFieldParam}
                        \membrPhaseFieldSym
                        +
                        \pfSpontMeanCurv
                    \right)
                    \left(
                        \membrPhaseFieldSym^2-1
                    \right)
                \right)^2
            }{\lebesgueM{3}}.
        $$
        So the ``diffuse'' Canham--Helfrich energy we employ in the following is 
        given by 
        $$
            \helfrichPFEnergyFunct{\membrPhaseFieldSym}{}
            =
            \pfGenWillmoreEnergyFunct{\membrPhaseFieldSym}{\pfSpontMeanCurv}
            +
            \ginzburgLandauEnergyFunct{\membrPhaseFieldSym}{}
        $$
        for the membrane phase field and analogously for the cortex phase field.
\subsection{Mechanics of the Bulk}       
    The bulk region $ \domain $ is assumed to be filled completely by a 
    fluid. The movement of the fluid body in the course of time shall be described by
    a sufficiently smooth invertible map $ \fluidBodyTransf{}{} $. At the 
    boundary $ \boundary{\domain} $ of $ \domain $,
    we prescribe that the wall is impenetrable by the fluid
    $$
        \pDiff{t}{}{}\fluidBodyTransf{}{x}\cdot\normal{} = 0,\quad \forall x \in \boundary{\domain},
    $$
    as well as no slip conditions
    $$
        \pDiff{t}{}{}\fluidBodyTransf{}{x}\cdot\tau = 0,\quad \forall x \in \boundary{\domain},
    $$
    where $ \tau $ is any tangential vector of $ \boundary{\domain} $.
    A fluid particle can be specified by its initial position (at $ t = 0 $)
    $ \bar{x} \in \domain $, also called its label or \emph{Lagrangian
    coordinate}, and its current position, or \emph{Eulerian coordinate},
    $ x = \fluidBodyTransf{t}{\bar{x}} $.
    The \emph{Lagrangian velocity} is
    given by $ \shapeTransfVel(t,\bar{x}) = \pDiff{t}{}{}\fluidBodyTransf{t}{\bar{x}} $, 
    and thus its
    Eulerian counterpart is $ \velocity{t}{x} = \shapeTransfVel(t,
    \invFluidBodyTransf{t}{x}) $. In this regard, we introduced two
    interpretation frames in which the velocity of a deformation may be studied: In the 
    Lagrangian frame $ \shapeTransfVel(t,x) $ assigns to a particle $ x \in \domain $ its velocity at 
    a given time. In the Eulerian frame, $ x $ is no longer
    a particle identifier, but a point in space to which the velocity $ \velocity{t}{x} $
    of the particle passing through $ x $ at time $t $ is assigned.
    In the following, we may switch between these
    two observer positions as suited.

All forces inside the Eulerian fluid body are defined by the \emph{Cauchy stress tensor}
$$
    \cauchyStressTens{}{} 
    = 
    \viscousStressTens{}{}
    -
    \pressure{}{}\identityMatrix[3],
$$
which can be separated into a part accounting for stress due to
particle friction, the \emph{viscous stress tensor} $ \viscousStressTens{}{} $,
and a part that captures the elastic collisions of the fluid particles
and is given by the \emph{hydrostatic pressure} $ \pressure{}{} $.

The fluid considered in our model 
shall be isotropic and Newtonian. For the viscous stress tensor, we assume
\begin{equation*}
    \viscousStressTens{}{}
    =
    \viscosity
    \left(
        \grad{}{}{}\velocity{}{}
        +
        \transposed{
            \grad{}{}{}\velocity{}{}                  	
        }
        -
        \frac{2}{3}
        \diver{}{}{}\velocity{}{}
        \identityMatrix
    \right)
    +
    \zeta
    \diver{}{}{}\velocity{}{}
\end{equation*}
(cf.~\cite[Chapter~II]{Landau+1987}),
where $ \viscosity $ and $ \zeta $ are coefficients describing
the viscosity of the fluid. This constitutive law prescribes a linearized version
of the viscous stress tensor and is to be employed with care when large velocity 
gradients appear. 

Furthermore, the fluid we consider throughout this work is incompressible, i.e.,
$$
    \diver{}{}{}\velocity{}{} = 0.
$$
A direct consequence is a simplification of the viscous stress tensor to
\begin{equation*}
    \viscousStressTens{}{}
    =
    \viscosity
    \left(
      \grad{}{}{}\velocity{}{}
      +
      \transposed{
        \grad{}{}{}\velocity{}{}                  	
      }
    \right).
\end{equation*}

\subsection{Membrane-Cortex Coupling}
    As pointed out above, linker proteins pin the membrane to the cortex. 
    In our model, we  consider them as \mbox{(macro-)}molecules sticking to the cortex thus being transported with it.
    In this situation, it is not clear how to model the direction
    with which these molecules point out of the cortex and connect to the membrane.
    To mitigate this problem and leave a certain degree of freedom, we 
    introduce a \emph{connectivity function} 
    $ \couplingProb{}{} $ such that
    $ \couplingProb{x,y,\normal[\cortexPhaseFieldSym]{}}{} $ 
    describes the probability with which
    a point $ x \in\reals^3$ on the cortex is connected to a point $ y\in\reals^3 $ on the membrane 
    in direction $ y-x $. The function also depends on the normal approximation 
    $ \normal[\cortexPhaseFieldSym]{x} = \frac{\grad{}{}{}\cortexPhaseFieldSym(x)
    }{\abs{\grad{}{}{}\cortexPhaseFieldSym(x)}} $ for all points $ x $ with
    $ \grad{}{}{}\cortexPhaseFieldSym(x) \neq 0 $,
    which could be considered a gauge direction.
    For example, we may choose
    \begin{equation}
        \label{equ:angle dependent connectivity}
        \couplingProb{x,y,\normal[\cortexPhaseFieldSym]{}}{}
        =
        \couplingProb[\tilde]{\frac{\normal[\cortexPhaseFieldSym]{x}\cdot(y-x)}{\abs{y-x}}}{}
    \end{equation}
    and
    \begin{equation}
        \label{equ:Gaussian direction distribution}
        \couplingProb[\tilde]{t}{} = \couplingProb[\hat]{}{}\exp\left(
            \frac{\arccos(t)^2}{s^2}
        \right)
    \end{equation}
    for some standard deviation $ s $ and a scaling factor $ \couplingProb[\hat]{}{} $
    to model Gaussian distribution of directions around the cortex normal.

    In order to define the potential energy in the system due to this coupling,
    we use the Hookean spring model for the linker proteins: 
    The energy assigned to a linker connecting $ x $ and $ y $ is 
    $$ 
        \frac{1}{2}\linkersSpringConst \abs{y-x}^2,
    $$
    where $ \linkersSpringConst $
    is a spring constant. 
    Let us assume the particle volume density of linkers connecting membrane
    and cortex at $ p $ is given
    by the function $ \activeLinkers{\NOARG}{x} $.
    In case every linker at $ x $ connects to $ y $,
    the energy volume density at $ x $ is then given by the expression
    $$
        \activeLinkers{\NOARG}{x} \frac{1}{2}\linkersSpringConst \abs{y-x}^2.
    $$
    However, not every linker at $ x $ might connect to $ y $. This is exactly,
    where $ \couplingProb{}{} $ comes into play: In the more general scenario
    we consider here, the energy volume density at $ x $ is given by 
    $$
        \integral{\domain}{}{
            \pfDiracApprox{\phaseFieldParam}{\membrPhaseFieldSym}(y)
            \couplingProb{x,y,\normal[\cortexPhaseFieldSym]{}}{}
            \activeLinkers{\NOARG}{p} 
            \frac{1}{2}\linkersSpringConst \abs{y-x}^2
        }{\lebesgueM{3}(y)},
    $$
    where the integral operator
    $$
        u \mapsto 
        \integral{\domain}{}{
            \pfDiracApprox{\phaseFieldParam}{\membrPhaseFieldSym}(y)
            u(y)
        }{\lebesgueM{3}(y)}
    $$
    is meant to integrate a quantity $ u $ over the membrane: Since we 
    only have phase fields and no two-dimensional surfaces, 
    we use a Dirac-like weight 
    $ \pfDiracApprox{\phaseFieldParam}{\membrPhaseFieldSym} $ to concentrate the 
    integral in the region of the diffuse membrane layer. We will describe below,
    how to choose $ \pfDiracApprox{\phaseFieldParam}{\membrPhaseFieldSym} $.
    Finally, the potential energy contained in the system due to this coupling 
    can be measured by the double-integral
    $$
        \pfCouplingEnergy{\membrPhaseFieldSym,\cortexPhaseFieldSym,\activeLinkersSym}{}
        \colonequals
        \integral{\domain}{}{
            \pfDiracApprox{\phaseFieldParam}{\cortexPhaseFieldSym}(x)
            \integral{\domain}{}{
                \pfDiracApprox{\phaseFieldParam}{\membrPhaseFieldSym}(y)
                \couplingProb{x,y,\normal[\cortexPhaseFieldSym]{}}{}
                \activeLinkers{\NOARG}{x} 
                \frac{1}{2}\linkersSpringConst \abs{y-x}^2
            }{\lebesgueM{3}(y)}
        }{\lebesgueM{3}(x)}.
    $$



    \subsection{Linker Protein Dynamics}
        Above we already used the density $ \activeLinkers{}{} $ of proteins
        that connect membrane and cortex; we call them \emph{active}.  
        We already mentioned that the proteins
        are ripped apart when overstretched, but it has also been reported \cite{CharrasPaluch2008}
        that these broken entities may be repaired (or `healed'). As we aim at 
        modeling the ripping and repairing processes, we introduce 
        further a density $ \inactiveLinkers{}{} $ of broken proteins that no longer connect,
        but may be repaired; these proteins are called \emph{inactive}.

        Both linker protein densities evolve due to three effects: 
        The fluid transporting the cortex and the particles on it, 
        surface diffusion, as well as ripping and repairing.
        If the cortex were a sharp interface, i.e., a surface, 
        we would consider the reaction-diffusion system
        \begin{subequations}
            \begin{align*}
                \matDeriv{t}{}{}\activeLinkers{}{}
                -
                \diver{\interfTwo{}}{}{
                    \speciesOneDiffusiv{}
                    \grad{}{}{}\speciesOne{}{}
                }
                +
                \activeLinkers{}{}
                \diver{\interfTwo{}}{}{}
                \pfVelocity{}{}
                &=
                \repairRate \inactiveLinkers{}{}
                -
                \discRate{\interfOne{}}
                \activeLinkers{}{}
                \quad\quad&\text{on}&\;\interfTwo{},
                \\
                \matDeriv{t}{}{}\inactiveLinkers{}{}
                -
                \diver{\interfTwo{}}{}{
                    \speciesTwoDiffusiv{}
                    \grad{}{}{}\speciesTwo{}{}
                }
                +
                \inactiveLinkers{}{}
                \diver{\interfTwo{}}{}{}
                \pfVelocity{}{}
                &=
                -\repairRate\inactiveLinkers{}{}
                +
                \discRate{\interfOne{}}
                \activeLinkers{}{}
                \quad\quad&\text{on}&\;\interfTwo{}.
            \end{align*}
        \end{subequations}
        The parameters of this system are as follows:
        $ \repairRate $ is a repairing rate, $ \discRate{\interfOne{}} $ a
        ripping rate that depends on the position of the membrane (if the membrane is
        far away from the cortex, the stretching of linkers is high; if the membrane is close,
        the stretching is low), and $ \speciesOneDiffusiv{}, \speciesTwoDiffusiv{}
        \in \reals $ are diffusivities. Note that the right hand sides have opposite signs, which
        ensures conservation of the number of particles on the surface.

        Now let us extend the densities $ \activeLinkers{}{} $, $ \inactiveLinkers{}{} $
        constantly in normal direction of $ \interfTwo{} $.
        We then can express
        $$
            \matDeriv{t}{}{}\rho_\alpha = \pDiff{t}{}{}\rho_\alpha
            +
            \pfVelocity{}{}
            \cdot
            \grad{}{}{}\rho_\alpha,
            \quad\quad \alpha \in \{a,i\}.
        $$
        Decomposing $ \velocity{}{} = \velocityNormal{}{}{}\normal[\interfTwo{}]{}
        + \symVelocityTangVec $ into a normal and tangential part, 
        we find 
        $$
            \pfVelocity{}{}
            \cdot
            \grad{}{}{}\rho_\alpha
            =
            \symVelocityTangVec
            \cdot
            \grad{}{}{}\rho_\alpha
        $$
        (the normal derivative of $\rho_\alpha $ is zero) and 
        $ \rho_\alpha \diver{\interfTwo{}}{}{}\pfVelocity{}{}
        =
        -\rho_\alpha \meanCurv{}{}\velocityNormal{}{}{}
        +
        \rho_\alpha \diver{\interfTwo{}}{}{}\symVelocityTangVec 
        $.
        Inserting into the above equations, we finally obtain
        \begin{subequations}
            \begin{align*}
                \pDiff{t}{}{}\activeLinkers{}{}
                -
                \meanCurv{}{}
                \symVelocityNormal{}{}{}
                \activeLinkers{}{}
                -
                \diver{\interfTwo{}}{}{
                    \speciesOneDiffusiv{}
                    \grad{}{}{}\speciesOne{}{}
                }
                +
                \diver{\interfTwo{}}{}{
                    \activeLinkers{}{}
                    \symVelocityTangVec
                }
                &=
                \repairRate \inactiveLinkers{}{}
                -
                \discRate{\interfOne{}}
                \activeLinkers{}{}
                \quad\quad&\text{on}&\;\interfTwo{},
                \\
                \pDiff{t}{}{}\inactiveLinkers{}{}
                -
                \meanCurv{}{}
                \symVelocityNormal{}{}
                \inactiveLinkers{}{}
                -
                \diver{\interfTwo{}}{}{
                    \speciesTwoDiffusiv{}
                    \grad{}{}{}\speciesTwo{}{}
                }
                +
                \diver{\interfTwo{}}{}{
                \inactiveLinkers{}{}
                \symVelocityTangVec
                }
                &=
                -\repairRate\inactiveLinkers{}{}
                +
                \discRate{\interfOne{}}
                \activeLinkers{}{}
                \quad\quad&\text{on}&\;\interfTwo{}.
            \end{align*}
        \end{subequations}

        Since we are not dealing with surfaces, but diffuse layers, we need to 
        approximate the surface differential operators.
        As a first step, we reformulate the above equations 
        in the sense of distributions employing 
        a Dirac distribution $ \delta_{\interfTwo{}} $ concentrating mass on $ \interfTwo{} $,
        and an extension $ \bar{\pfVelocity{}{}} $ of the velocity
        $ \pfVelocity{}{} $ with $ \normal[\interfTwo{}]{}\tensorProd
        \normal[\interfTwo{}]{} \frobProd \grad{}{}{}\pfVelocity[\bar]{}{} = 0 $ such 
        that $ \diver{}{}{}\bar{\pfVelocity{}{}} = \diver{\interfTwo{}}{}{}\pfVelocity{}{} $.
        The reformulated equations contain
        only bulk differential operators:
        \begin{subequations}
            \begin{align*}
                \delta_{\interfTwo{}}
                \pDiff{t}{}{}\activeLinkers{}{}
                -
                \delta_{\interfTwo{}}
                \meanCurv{}{}
                \bar{\symVelocityNormal{}{}{}}
                \activeLinkers{}{}
                -
                \diver{}{}{
                    \delta_{\interfTwo{}}
                    \speciesOneDiffusiv{}
                    \grad{}{}{}\speciesOne{}{}
                }
                +
                \diver{}{}{
                    \delta_{\interfTwo{}}
                    \activeLinkers{}{}
                    \bar{\pfVelocity{}{}}_\tau
                }
                &=
                \delta_{\interfTwo{}}(
                \repairRate \inactiveLinkers{}{}
                -
                \discRate{\membrPhaseFieldSym}
                \activeLinkers{}{})
                \quad\quad&\text{on}&\;\interfTwo{},
                \\
                \delta_{\interfTwo{}}
                \pDiff{t}{}{}\inactiveLinkers{}{}
                -
                \delta_{\interfTwo{}}
                \meanCurv{}{}
                \bar{\symVelocityNormal{}{}}
                \inactiveLinkers{}{}
                -
                \diver{}{}{
                    \delta_{\interfTwo{}}
                    \speciesTwoDiffusiv{}
                    \grad{}{}{}\speciesTwo{}{}
                }
                +
                \diver{}{}{
                    \delta_{\interfTwo{}}
                    \inactiveLinkers{}{}
                    \bar{\pfVelocity{}{}}_\tau
                }
                &=
                -\delta_{\interfTwo{}}
                (
                \repairRate\inactiveLinkers{}{}
                +
                \discRate{\membrPhaseFieldSym}
                \activeLinkers{}{}
                )
                \quad\quad&\text{on}&\;\interfTwo{}.
            \end{align*}
        \end{subequations}

        We then approximate $ \delta_{\interfTwo{}} $ by a smooth function
        $ \pfDiracApprox{\phaseFieldParam}{\cortexPhaseField{}{}} $ that
        converges to $ \delta_{\interfTwo{}} $ for $ \phaseFieldParam \to 0 $ in the sense
        of distributions. Also, we introduce phase field analogues for the mean curvature,
        the normal part of the velocity, and the corresponding tangential part, respectively:
        \begin{equation*}
            \begin{split}
                \meanCurv{\symCortexPhaseField}{}
                =
                \abs{\grad{}{}{}\symCortexPhaseField}
                \left(
                    -
                    \phaseFieldParam
                    \laplacian{}{}{}\symCortexPhaseField
                    +
                    \phaseFieldParam^{-1}
                    \doubleWellPot[\schemeSupscr{\prime}]{\symCortexPhaseField}
                \right),\quad\quad
                \symVelocityNormal{\membrPhaseFieldSym}{} = (\pfVelocity{}{} \cdot 
                \normal[\cortexPhaseField{}{}]{})\normal[\cortexPhaseField{}{}]{}
                + (\pfVelocity{}{})_\tau.         
            \end{split}
        \end{equation*}
        Thus, the following system in the bulk $\Omega$ is obtained:
        \begin{subequations}
            \label{equ:modelling:phase field model:species subsystem}
            \begin{align}
                \pfDiracApprox{\phaseFieldParam}{\cortexPhaseField{}{}}
                \pDiff{t}{}{} \speciesOne{}{} 
                -
                \pfDiracApprox{\phaseFieldParam}{\cortexPhaseField{}{}}                
                \meanCurv{}{}
                \symVelocityNormal{\cortexPhaseFieldSym}{}
                \pfSpeciesOne{}{}
                -
                \diver{}{}{
                    \pfDiracApprox{\phaseFieldParam}{\cortexPhaseField{}{}}
                    \speciesOneDiffusiv{}
                    \grad{}{}{}\speciesOne{}{}
                }
                +
                \diver{}{}{
                    \pfDiracApprox{\phaseFieldParam}{\cortexPhaseField{}{}}                
                    (\pfVelocity{}{})_\tau
                    \pfSpeciesOne{}{}
                }
                &=
                \pfDiracApprox{\phaseFieldParam}{\cortexPhaseField{}{}}
                (
                \repairRate\inactiveLinkers{}{}
                -
                \discRate{\membrPhaseFieldSym}
                \activeLinkers{}{}
                ),&
                \\
                \pfDiracApprox{\phaseFieldParam}{\cortexPhaseField{}{}}
                \pDiff{t}{}{}\speciesTwo{}{} 
                -
                \pfDiracApprox{\phaseFieldParam}{\cortexPhaseField{}{}}                
                \meanCurv{}{}
                \symVelocityNormal{\cortexPhaseFieldSym}{}
                \pfSpeciesTwo{}{}
                -
                \diver{}{}{
                    \pfDiracApprox{\phaseFieldParam}{\cortexPhaseField{}{}}
                    \speciesTwoDiffusiv{}
                    \grad{}{}{}\speciesTwo{}{}
                }
                +
                \diver{}{}{
                    \pfDiracApprox{\phaseFieldParam}{\cortexPhaseField{}{}}                
                    (\pfVelocity{}{})_\tau
                    \pfSpeciesTwo{}{}
                }
                &=
                \pfDiracApprox{\phaseFieldParam}{\cortexPhaseField{}{}}
                (
                -\repairRate\inactiveLinkers{}{}
                +
                \discRate{\membrPhaseFieldSym}
                \activeLinkers{}{}
                ).&                
            \end{align}
        \end{subequations}

        Now that the structure of the linker dynamics is fixed, it remains to specify the function $ \discRate{\membrPhaseFieldSym} $
        giving the ripping rate. Let us first discuss this function in the sharp interface 
        setting. To obtain the ripping rate, we integrate over the membrane, where we 
        weigh the area element by the connectivity and a ripping density $ \tilde{\discRate{}} $
        $$
            \discRate{\interfOne{}} = \integral{\interfOne{}}{}{
                \couplingProb{x,y,\normal[\interfTwo{}]{}}{}
                \tilde{\discRate{}}(x,y)
            }{\hausdorffM{2}(y)}.
        $$
        The most common models for the ripping density are on the one hand a Kramer-type 
        kinetic model derived in \cite{Evans2001}, and also employed by \cite{Alert+2015}:
        \begin{equation}
            \label{equ:ripping density model:evans}
            \tilde{\discRate{}}(x,y) =
            \koff 
            \exp\left(
              \frac{ \linkersSpringConst{} \abs{y-x} \charEnergyLen }{ \temperature \boltzmConst } 
            \right).
        \end{equation}
        On the other hand, \cite{Lim+2012} suggested a discontinuous approach towards
        modelling the linker forces: They take the linker force as the product of the 
        Hookean force term and a step function:
        $$ \linkersSpringConst{}\abs{y-x} \left(1-\frac{\abs{y-x }}{\critLen}\right)^+ $$
        with $ (\cdot)^+ $ denoting the non-negative part and  $ \ell^* $ a 
        critical length above which the linkers rip.
        In this model, the linker ripping rate can be interpreted as being zero below the
        critical height and infinity above. Motivated by this approach, 
        a ripping density of
        $$
            \tilde{\discRate{}}(x,y) = \frac{(\abs{y-x}-\ell^*)^+}{\discSteepness}
        $$
        for $ \discSteepness \ll 1 $ has been used in \cite{Werner+2020}
        as a continuous interpolation.
        Going from sharp to diffuse, we only need to introduce a weight under the integral
        $$
            \integral{\interfOne{}}{}{
                \couplingProb{x,y,\normal[\interfTwo{}]{}}{}
                \tilde{\discRate{}}(x,y)
            }{\hausdorffM{2}(y)}
            \longrightarrow
            \integral{\domain}{}{
                \pfDiracApprox{\phaseFieldParam}{\membrPhaseFieldSym}
                \couplingProb{x,y,\normal[\cortexPhaseFieldSym]{}}{}
                \tilde{\discRate{}}(x,y)
            }{\lebesgueM{3}(y)}.
        $$

\section{A PDE Model via Onsager's Principle}
    From these considerations, we can make an ansatz for a system of partial differential equations
    (PDEs) that should describe our physical system:
    \begin{subequations}
        \label{equ:onsagers principle:ansatz}
        \begin{align}
            \label{equ:onsagers principle:momentum conservation} 
            \fluidMassDens(
            \pDiff{t}{}{}\pfVelocity{}{} 
            +
            \grad{}{}{}
            \pfVelocity{}{}
            \pfVelocity{}{})
            -
            \diver{}{}{}\symVelocityFlux &= K,
            \\
            \diver{}{}{}\pfVelocity{}{} &= 0,
            \\
            \label{equ:onsagers principle:membrane cont equ} 
            \pDiff{t}{}{}\symMembrPhaseField 
            +
            \symVelocity
            \cdot
            \grad{}{}{}
            \symMembrPhaseField
            - 
            \diver{}{}{}\symMembrPhaseFieldFlux &= 0,
            \\
            \label{equ:onsagers principle:cortex cont equ}
            \pDiff{t}{}{}\symCortexPhaseField 
            +
            \symVelocity
            \cdot
            \grad{}{}{}
            \symCortexPhaseField
            - 
            \diver{}{}{}\symCortexPhaseFieldFlux &= 0,
            \\
            \label{equ:onsagers principle:species one cont equ}
            \pfDiracApprox{\phaseFieldParam}{\cortexPhaseField{}{}}
            \pDiff{t}{}{}\pfSpeciesOne{}{}
            -
            \symVelocityNormal{\cortexPhaseField{}{}}
            \meanCurv{\cortexPhaseField{}{}}{}
            \pfSpeciesOne{}{}
            -
            \diver{}{}{
                \pfDiracApprox{\phaseFieldParam}{\cortexPhaseField{}{}}
                \speciesOneDiffusiv{}
                \grad{}{}{}\pfSpeciesOne{}{}
            }
            +
            \diver{}{}{
                \pfDiracApprox{\phaseFieldParam}{\cortexPhaseField{}{}}
                \pfSpeciesOne{}{}
                (\pfVelocity{}{})_\tau
            }
            &=\\\nonumber
            \pfDiracApprox{\phaseFieldParam}{\cortexPhaseField{}{}}
            (
            \repairRate\inactiveLinkers{}{}
            -
            \discRate{\membrPhaseFieldSym}
            \activeLinkers{}{}
            )&,
            \\
            \label{equ:onsagers principle:species two cont equ}
            \pfDiracApprox{\phaseFieldParam}{\cortexPhaseField{}{}}
            \pDiff{t}{}{}\pfSpeciesTwo{}{}
            -
            \symVelocityNormal{\cortexPhaseField{}{}}
            \meanCurv{\cortexPhaseField{}{}}{}
            \pfSpeciesTwo{}{}
            -
            \diver{}{}{
                \pfDiracApprox{\phaseFieldParam}{\cortexPhaseField{}{}}
                \speciesTwoDiffusiv{}
                \grad{}{}{}\pfSpeciesTwo{}{}
            }
            +
            \diver{}{}{
                \pfDiracApprox{\phaseFieldParam}{\cortexPhaseField{}{}}
                \pfSpeciesTwo{}{}
                (\pfVelocity{}{})_\tau
            }
            &=\\\nonumber
            \pfDiracApprox{\phaseFieldParam}{\cortexPhaseField{}{}}
            (
            -\repairRate\inactiveLinkers{}{}
            +
            \discRate{\membrPhaseFieldSym}
            \activeLinkers{}{}
            )&,
            \\
            \label{equ:modelling:phase field:phase fields Neumann conditions}
            \restrFun{
                \pDiff{\normal{}}{}{}\membrPhaseField{}{}
            }{
                \boundary{\domain}
            }
            =
            \restrFun{
                \pDiff{\normal{}}{}{}\cortexPhaseField{}{}
            }{
                \boundary{\domain}
            }
            &=
            0,
            \\
            \label{equ:modelling:phase field:flux condition}
            \restrFun{\symMembrPhaseFieldFlux}{\boundary{\domain}}\cdot\normal{} =
            \restrFun{\symCortexPhaseFieldFlux}{\boundary{\domain}}\cdot\normal{} &= 0,
            \\
            \label{equ:modelling:phase field:species one Dirichlet condition}
            \restrFun{\pfSpeciesOne{}{}}{\boundary{\domain}} &= 0,
            \\
            \label{equ:modelling:phase field:species two Dirichlet condition}
            \restrFun{\pfSpeciesTwo{}{}}{\boundary{\domain}} &= 0,
        \end{align}
    \end{subequations}
    and search for the thermodynamic fluxes $ \symVelocityFlux $, $ \symMembrPhaseFieldFlux $,
    $ \symCortexPhaseFieldFlux $ and the force $ K $.
    This is done by employing Onsager's principle that 
    postulates 
    \begin{equation}
        \label{equ:modelling:phase field model:Onsager optimisation problem}
        \min_{\symVelocityFlux,\symMembrPhaseFieldFlux,\symCortexPhaseFieldFlux}
        \left(
            \diff{t}{}{
                \totalEnergyFunct{
                    \pfVelocity{}{},
                    \membrPhaseFieldSym,
                    \cortexPhaseFieldSym,
                    \pfSpeciesOne{}{}
                }
            }
            +
            \dissipFunct{
                \symVelocityFlux,
                \symMembrPhaseFieldFlux,
                \symCortexPhaseFieldFlux
            }{}
        \right),
    \end{equation}
    where the internal energy is given as the sum of the kinetic and 
    free energy in the system:
    $$ 
        \totalEnergyFunct{
            \pfVelocity{}{},
            \membrPhaseFieldSym,
            \cortexPhaseFieldSym,
            \pfSpeciesOne{}{}
        }
        =
        \frac{1}{2}\LTNorm{\domain}{\pfVelocity{}{}}^2
        +
        \freeEnergyFunct{
            \membrPhaseFieldSym,
            \cortexPhaseFieldSym,
            \pfSpeciesOne{}{}
        }{},
    $$
    and 
    \begin{equation*}
        \dissipFunct{\symVelocityFlux,\symMembrPhaseFieldFlux,\symCortexPhaseFieldFlux}{} 
        = 
        \integral{\domain}{}{
            \frac{\abs{\symVelocityFlux}^2}{2\viscosity}
            +
            \frac{\abs{\symMembrPhaseFieldFlux}^2}{2\membrPhaseFieldMob{\symMembrPhaseField}}
            +
            \frac{\abs{\symCortexPhaseFieldFlux}^2}{
                2\cortexPhaseFieldMob{\symCortexPhaseField}
            }  
        }{\lebesgueM{3}}
    \end{equation*}
    is the functional that gives the dissipation in the system for the 
    mobilities $ \membrPhaseFieldMob{\membrPhaseFieldSym} $, 
    $ \cortexPhaseFieldMob{\cortexPhaseFieldSym} $.

    We obviously need to calculate the time derivative of the internal energy.
    Before we will do so, we shall elaborate on the choice of the ``diffuse Dirac 
    functions'' $ \pfDiracApprox{\phaseFieldParam}{\membrPhaseFieldSym} $
    and $ \pfDiracApprox{\phaseFieldParam}{\cortexPhaseFieldSym} $.
    \subsection{Phase Field Transport Formula}
        In the sharp interface setting, we encounter surface integrals
        $ \integral{\fluidBodyTransf{t}{\interfTwo{}}}{}{f}{\lebesgueM{3}} $,
        for $ \funSig{f}{\reals\times\domain}{\reals} $,
        whose time derivatives are
        treated by leveraging the classical surface transport formula
        (cf., e.g., \cite{Pruess+2016})
        \begin{equation*}
            \diff[t^*]{t}{}{
                \integral{\fluidBodyTransf{t}{\interfTwo{}}}{}{f}{\lebesgueM{3}}
            }
            =
            \integral{\fluidBodyTransf{t^*}{\interfTwo{}}}{}{
                \matDeriv[t^*]{t}{f}
                +
                f
                \diver{\fluidBodyTransf{t^*}{\interfTwo{}}}{}{}
                \pDiff{t}{}{}\fluidBodyTransf{}{}
            }{\hausdorffM{2}}.
        \end{equation*}
        From the phase field perspective, surface integrals make no sense since there is no
        such two-dimensional structure. Nonetheless, the terms produced by application
        of the transport formula need to be represented in the phase field PDE system
        in order to achieve consistency with the corresponding sharp 
        interface system for $ \phaseFieldParam \searrow 0 $.
        Let us consider the bulk integral 
        $  \integral{\fluidBodyTransf{t}{\domain}}{}{
                f \pfDiracApprox{\phaseFieldParam}{\membrPhaseField{}{}}
           }{\lebesgueM{3}} 
        $
        weighted by a characteristic for the diffuse layer analogue of $ \interfTwo{} $.
        The fluid deformation $ \fluidBodyTransf{}{} $ maps $ \domain $ onto itself, so 
        $$ 
            \diff{t}{}{
                \integral{\fluidBodyTransf{t}{\domain}}{}{
                    f \pfDiracApprox{\phaseFieldParam}{\membrPhaseField{}{}}
                }{\lebesgueM{3}}
            } 
            =
            \integral{\domain}{}{
                \pDiff{t}{}{
                    f \pfDiracApprox{\phaseFieldParam}{\membrPhaseField{}{}}
                }
            }{\lebesgueM{3}}
            =
            \integral{\domain}{}{
                \pfDiracApprox{\phaseFieldParam}{\membrPhaseField{}{}}              	
                \pDiff{t}{}{}f
                +
                f
                \pDiff{t}{}{
                   \pfDiracApprox{\phaseFieldParam}{\membrPhaseField{}{}}
                }
            }{\lebesgueM{3}}.
        $$
        To achieve the transport terms, the choice of $ \pfDiracApprox{\phaseFieldParam}{\membrPhaseField{}{}} $
        is critical. 
        Let us consider the choice $ \pfDiracApprox{\phaseFieldParam}{\membrPhaseField{}{}}
        = \ginzburgLandauEnergyDens{\membrPhaseField{}{}}{} $.
        The general form of a phase field equation is
        \begin{equation*}
            \pDiff{t}{}{}\symMembrPhaseField
            +
            \velocity{}{}
            \cdot
            \grad{}{}{}\symMembrPhaseField
            =
            \diver{}{}{}J(\membrPhaseField{}{}),
        \end{equation*}
        where $ \velocity{t}{x} = \pDiff{t}{}{}\fluidBodyTransf{t}{\fluidBodyTransf[\schemeSupscr{-1}]{t}{x}} $
        and
        $ J $ typically represents a thermodynamical flux. If $ \grad{}{}{}\membrPhaseField{}{} \neq 0 $,
        we may define 
        $ \normal[\membrPhaseField{}{}]{} = 
        \frac{\grad{}{}{}\membrPhaseField{}{}}{\abs{\grad{}{}{}\membrPhaseField{}{}}} $ and decompose
        $ 
            \velocity{}{} = 
            \symVelocityNormalVec{\membrPhaseField{}{}}
            + 
            \symVelocityTangVec{}
            =
            \symVelocityNormal{\membrPhaseField{}{}} \normal[\membrPhaseField{}{}]{} 
            + 
            \symVelocityTangVec{}.
        $

        Taking the time-derivative of the integral
        $
            \integral{\domain}{}{
                \ginzburgLandauEnergyDens{\symMembrPhaseField}{}
                f
            }{\lebesgueM{3}},
        $
        we see
        \begin{equation*}
            \begin{split}
                \diff{t}{}{}
                \integral{\domain}{}{
                    \ginzburgLandauEnergyDens{\symMembrPhaseField}{}
                    f
                }{\lebesgueM{3}}
                &=
                \integral{\domain}{}{
                    \ginzburgLandauEnergyDens{\symMembrPhaseField}{}
                    \pDiff{t}{}{}f
                }{\lebesgueM{3}}
                +
                \integral{\domain}{}{
                    \gatDeriv{\symMembrPhaseField}{}{\ginzburgLandauEnergyDens{}{}}{%
                        \pDiff{t}{}{}\symMembrPhaseField
                    }
                    f
                }{\lebesgueM{3}}
                ,
            \end{split}
        \end{equation*}
        where
        \begin{equation*}
            \begin{split}
                \gatDeriv{\symMembrPhaseField}{}{
                    \ginzburgLandauEnergyDens{}{}
                }{\pDiff{t}{}{}\symMembrPhaseField}
                &=
                \phaseFieldParam
                \grad{}{}{}\symMembrPhaseField
                \cdot
                \grad{}{}{\pDiff{t}{}{}\symMembrPhaseField}
                +
                \phaseFieldParam^{-1}
                \doubleWellPot[\schemeSupscr{\prime}]{\symMembrPhaseField}
                \pDiff{t}{}{}\symMembrPhaseField.
            \end{split}
        \end{equation*}
        Furthermore, using  integration by parts and the boundary condition
            $ \restrFun{\pDiff{\normal{}}{}{}\membrPhaseField{}{}}{\boundary{\domain}} = 0 $ we obtain
        \begin{equation}
            \label{equ:modelling:phase field model:diffuse layer transport formula}
            \begin{split}
                \integral{\domain}{}{
                    \gatDeriv{\symMembrPhaseField}{}{\ginzburgLandauEnergyDens{}{}}{%
                        \pDiff{t}{}{}\symMembrPhaseField
                    }
                    f
                }{\lebesgueM{3}}
                &=
                \integral{\domain}{}{
                    \phaseFieldParam
                    \grad{}{}{}\symMembrPhaseField
                    \cdot
                    \grad{}{}{\pDiff{t}{}{}\symMembrPhaseField}
                    f
                    +
                    \phaseFieldParam^{-1}
                    \doubleWellPot[\schemeSupscr{\prime}]{\symMembrPhaseField}
                    \pDiff{t}{}{}\symMembrPhaseField
                    f
                }{\lebesgueM{3}}
                \\
                &=
                -
                \integral{\domain}{}{
                    \phaseFieldParam\laplacian{}{}{}\symMembrPhaseField
                    f
                    \pDiff{t}{}{}\symMembrPhaseField
                    +
                    \phaseFieldParam
                    \grad{}{}{}\symMembrPhaseField
                    \cdot
                    \grad{}{}{}f
                    \pDiff{t}{}{}\symMembrPhaseField
                }{\lebesgueM{3}}
                +
                \integral{\domain}{}{                
                    \phaseFieldParam^{-1}
                    \doubleWellPot[\schemeSupscr{\prime}]{\symMembrPhaseField}
                    \pDiff{t}{}{}\symMembrPhaseField
                    f
                }{\lebesgueM{3}}
                \\
                &=
                -
                \integral{\domain}{}{
                    \abs{\grad{}{}{}\symMembrPhaseField}
                    \symVelocityNormal{\membrPhaseField{}{}}
                    \left(
                        -
                        \phaseFieldParam
                        \laplacian{}{}{}\symMembrPhaseField
                        +
                        \phaseFieldParam^{-1}
                        \doubleWellPot[\schemeSupscr{\prime}]{\symMembrPhaseField}
                    \right)
                    f
                }{\lebesgueM{3}}
                +
                \integral{\domain}{}{
                    \phaseFieldParam
                    \abs{\grad{}{}{}\symMembrPhaseField}^2
                    \symVelocityNormalVec{\membrPhaseField{}{}} 
                    \cdot
                    \grad{}{}{}f
                }{\lebesgueM{3}}
                \\
                &+ \integral{\domain}{}{
                    f
                    \diver{}{}{}J(\membrPhaseField{}{})
                }{\lebesgueM{3}}
                \\
                &=
                -
                \integral{\domain}{}{
                    \symVelocityNormal{\membrPhaseField{}{}}
                    \meanCurv{\symMembrPhaseField}{}f
                }{\lebesgueM{3}}
                +
                \integral{\domain}{}{
                    \phaseFieldParam
                    \abs{\grad{}{}{}\symMembrPhaseField}^2
                    \symVelocityNormalVec{\membrPhaseField{}{}}\cdot\grad{}{}{}f
                }{\lebesgueM{3}}
                + 
                \integral{\domain}{}{
                    f
                    \diver{}{}{}J(\membrPhaseField{}{})
                }{\lebesgueM{3}}.
            \end{split}
        \end{equation}
        We set
        \begin{equation*}
            \begin{split}
                \meanCurv{\symMembrPhaseField}{}
                =
                \abs{\grad{}{}{}\symMembrPhaseField}
                \left(
                    -
                    \phaseFieldParam
                    \laplacian{}{}{}\symMembrPhaseField
                    +
                    \phaseFieldParam^{-1}
                    \doubleWellPot[\schemeSupscr{\prime}]{\symMembrPhaseField}
                \right)
            \end{split}
        \end{equation*}
        since for sigmoidal $ \symMembrPhaseField $ this is the
        mean curvature of level sets in a neighbourhood around $ \symMembrPhaseField^{-1}\{0\} $ 
        up to leading order in $\phaseFieldParam$.

        So all in all,
        \begin{equation*}
            \begin{split}
                \diff{t}{}{}
                \integral{\domain}{}{
                    \ginzburgLandauEnergyDens{\symMembrPhaseField}{}
                    f
                }{\lebesgueM{3}}
                &=
                -
                \integral{\domain}{}{
                    \symVelocityNormal{\membrPhaseField{}{}}
                    \meanCurv{\symMembrPhaseField}{}f
                }{\lebesgueM{3}}
                +
                \integral{\domain}{}{
                    \ginzburgLandauEnergyDens{\symMembrPhaseField}{}
                    \pDiff{t}{}{}f
                    +
                    \phaseFieldParam
                    \abs{\grad{}{}{}\symMembrPhaseField}^2
                    \symVelocityNormalVec{\membrPhaseField{}{}}
                    \cdot
                    \grad{}{}{}f
                }{\lebesgueM{3}}
                +
                \integral{\domain}{}{
                    f
                    \diver{}{}{}J(\membrPhaseField{}{})
                }{\lebesgueM{3}}
                \\
                &=
                -
                \integral{\symMembrPhaseField^{-1}\{0\}}{}{
                    \symVelocityNormal{\membrPhaseField{}{}}
                    \meanCurv{}{}f
                }{\lebesgueM{3}}
                +
                \integral{\symMembrPhaseField^{-1}\{0\}}{}{
                    \ginzburgLandauEnergyDens{\symMembrPhaseField}{}
                    \pDiff{t}{}{}f
                }{\lebesgueM{3}}        
                +
                \integral{\symMembrPhaseField^{-1}\{0\}}{}{
                    \symVelocityNormalVec{\membrPhaseField{}{}}\cdot\grad{}{}{}f
                }{\lebesgueM{3}}
                \\
                &+
                \integral{\domain}{}{
                    f
                    \diver{}{}{}J(\membrPhaseField{}{})
                }{\lebesgueM{3}}               
                +
                \landauBigO{\phaseFieldParam}
                \\
                &=
                \diff{t}{}{}
                \integral{\symMembrPhaseField^{-1}\{0\}}{}{
                    f
                }{\lebesgueM{3}}
                +
                \integral{\domain}{}{
                    f
                    \diver{}{}{}J(\membrPhaseField{}{})
                }{\lebesgueM{3}}
                +
                \landauBigO{\phaseFieldParam}.
            \end{split}
        \end{equation*}
        Neglecting the thermodynamical flux term, we may think of this as a phase field
        surface transport theorem.             	
        \paragraph{The Linker Proteins as Surfactants}
        With $ \pfDiracApprox{\membrPhaseFieldSym}{\phaseFieldParam}
        = \ginzburgLandauEnergyDens{\membrPhaseFieldSym}{} $
        the coupling energy can be interpreted as a surface
        energy of the membrane with a nonlinear elastic compression
        modulus
        $
            \couplingEnergyDens[\tilde]{y}{\speciesOne{}{},\normal[\cortexPhaseField{}{}]{}}
            =
        	\integral{\domain}{}{
              	\ginzburgLandauEnergyDens{\cortexPhaseFieldSym}{}(x)
	        	\couplingEnergyDens{x,y}{\speciesOne{}{},\normal[\cortexPhaseField{}{}]{}}
            }{\lebesgueM{3}(x)}.
        $
        Energies similar to
        $
            \integral{\domain}{}{
              	\ginzburgLandauEnergyDens{\membrPhaseFieldSym}{}(y)
                \couplingEnergyDens[\tilde]{y}{\speciesOne{}{},\normal[\cortexPhaseField{}{}]{}}
            }{\hausdorffM{2}(y)}            	
        $
        have been investigated in works on surfactants, see 
        \cite{Abels+2017, Garcke+2014, Aland+2017, Dunbar+2019}.
        Since the species influence the surface energy by their
        density $ \pfSpeciesOne{}{} $, they may also be regarded as
        surfactants.        

    \subsection{Stationarity Condition}
        To derive stationarity conditions for 
        \eqref{equ:modelling:phase field model:Onsager optimisation problem},
        we shall compute the time derivative of the internal energy
        $$ 
            \totalEnergyFunct{
                \pfVelocity{}{},
                \membrPhaseFieldSym,
                \cortexPhaseFieldSym,
                \pfSpeciesOne{}{}
            }{} 
            =
            \frac{1}{2}
            \integral{\domain}{}{
                    \abs{\pfVelocity{}{}}^2
            }{\lebesgueM{3}}
            +
            \helfrichPFEnergyFunct{\membrPhaseFieldSym}{}
            +
            \helfrichPFEnergyFunct{\cortexPhaseFieldSym}{}
            +
            \pfCouplingEnergy{\membrPhaseFieldSym,\cortexPhaseFieldSym,\pfSpeciesOne{}{}}{},
        $$
        First, let us
        split the velocity $ \symVelocity $
        into a normal and a tangential component with respect
        to the level sets of the phase fields:
        $
            \symVelocity 
            =
            \symVelocityNormalVec{\symMembrPhaseField}
            +
            \symVelocityTangVec{\symMembrPhaseField}
            =
            \symVelocityNormal{\symMembrPhaseField} \normal[\symMembrPhaseField]{}
            +
            \symVelocityTangVec{\symMembrPhaseField}.
        $
        Let us start with the kinetic energy part:
        \begin{equation*}
            \begin{split}
                \frac{1}{2}
                \diff{t}{}{}
                \integral{\domain}{}{
                    \abs{\velocity{}{}}^2
                }{\lebesgueM{3}}
                &=
                \integral{\domain}{}{
                    \symVelocity
                    \cdot
                    \pDiff{t}{}{}\symVelocity
                }{\lebesgueM{3}}
                \\
                &=
                \integral{\domain}{}{
                    \transposed{\symVelocity}
                    \left(
                        -\grad{}{}{}\symVelocity
                        \symVelocity
                        -
                        \diver{}{}{}\symVelocityFlux
                        +
                        K
                    \right)
                }{\lebesgueM{3}}
                \\
                &=
                \integral{\domain}{}{
                    \transposed{\symVelocity}
                    \left(
                        -
                        \diver{}{}{}\symVelocityFlux
                        +
                        K
                    \right)
                }{\lebesgueM{3}},
            \end{split}
        \end{equation*}
        where we inserted
        \eqref{equ:onsagers principle:momentum conservation} 
        and used the incompressibility of the fluid.
        For computing the time derivative of the coupling energy functional, we set
        \begin{align*}
            \pfCouplingIntCortex(t,y) &= \integral{\domain}{}{
                \ginzburgLandauEnergyDens{\symCortexPhaseField}{}(x)
                \pfCouplingEnergyDens{x,y}{\speciesOne{t}{x},
                  \normal[\cortexPhaseField{t}{\NOARG}]{x}}
            }{\lebesgueM{3}(x)},
            \\
            \pfCouplingIntMembr(t,x) &= \integral{\domain}{}{
                \ginzburgLandauEnergyDens{\membrPhaseFieldSym}{}(y)
                \pfCouplingEnergyDens{x,y}{\speciesOne{t}{x},
                  \normal[\cortexPhaseField{t}{\NOARG}]{x}}
            }{\lebesgueM{3}(y)},
        \end{align*}
        and then compute
        \begin{equation*}
            \begin{split}
                \diff{t}{}{
                    \pfCouplingEnergy{
                        \membrPhaseField{}{},
                        \cortexPhaseField{}{},
                        \pfSpeciesOne{}{}
                    }{}
                }
                &=
                \diff{t}{}{
                    \integral{\domain}{}{
                        \ginzburgLandauEnergyDens{\symMembrPhaseField}{}(y)
                        \pfCouplingIntCortex(t,y)
                    }{\lebesgueM{3}(y)}
                }
                \\
                &=
                \integral{\domain}{}{
                    \ginzburgLandauEnergyDens{\symMembrPhaseField}{}
                    \pDiff{t}{}{}\pfCouplingIntCortex
                    +
                    \phaseFieldParam
                    \abs{\grad{}{}{}\symMembrPhaseField}^2
                    \symVelocityNormalVec{\symMembrPhaseField}
                    \cdot
                    \grad{y}{}{}
                    \pfCouplingIntCortex
                    -
                    \symVelocityNormal{\symMembrPhaseField}
                    \meanCurv{\symMembrPhaseField}{}
                    \pfCouplingIntCortex
                    -
                    \pfCouplingIntCortex
                    \gatDeriv{\symMembrPhaseField}{}{
                        \ginzburgLandauEnergyDens{}{}
                    }{
                        \diver{}{}{}\symMembrPhaseFieldFlux
                    }
                }{\lebesgueM{3}}.
            \end{split}
        \end{equation*}
        For the last equality, we employ the phase fiel transport formula
        and \eqref{equ:onsagers principle:membrane cont equ}.
        We further deal with the time derivative:
        \begin{equation*}
            \begin{split}
                \pDiff{t}{}{}\pfCouplingIntCortex(t,y)
                &=
                \diff{t}{}{}
                \integral{\domain}{}{
                    \ginzburgLandauEnergyDens{\symCortexPhaseField}{}
                    (x)
                    \pfCouplingEnergyDens{x,y}{\pfSpeciesOne{}{},\normal[\cortexPhaseField{}{}]{}}
                }{\lebesgueM{3}(x)}
                \\
                &=
                \integral{\domain}{}{
                    \ginzburgLandauEnergyDens{\symCortexPhaseField}{}(x)
                    \diff{t}{}{
                        \pfCouplingEnergyDens{x,y}{
                            \pfSpeciesOne{}{},
                            \normal[\cortexPhaseField{}{}]{}
                        }
                    }
                    +
                    \phaseFieldParam
                    \abs{\grad{}{}{}\symCortexPhaseField}^2(x)
                    \symVelocityNormalVec{\symCortexPhaseField}(x)
                    \cdot
                    \grad{x}{}{
                    \pfCouplingEnergyDens{\cdot,y}{\pfSpeciesOne{}{},\normal[\cortexPhaseField{}{}]{}}
                    }
                }{\lebesgueM{3}(x)}
                \\
                &+
                \integral{\domain}{}{
                    -
                    \symVelocityNormal{\symCortexPhaseField}(x)
                    \meanCurv{\symCortexPhaseField}{x}
                    \pfCouplingEnergyDens{x,y}{\pfSpeciesOne{}{},\normal[\cortexPhaseField{}{}]{}}
                    -
                    \pfCouplingEnergyDens{x,y}{\speciesOne{}{},\normal[\cortexPhaseField{}{}]{}}
                    \diff{}{}{\ginzburgLandauEnergyDens{}{}}(
                        \diver{}{}{}\symCortexPhaseFieldFlux
                    )
                }{\lebesgueM{3}(x)}.
            \end{split}
        \end{equation*}
        Again, the phase field transport formula
        and 
        \eqref{equ:onsagers principle:cortex cont equ} are used.
        The term
        $                  	
            \diff{t}{}{\pfCouplingEnergyDens{x,y}{\pfSpeciesOne{}{},\normal[\cortexPhaseField{}{}]{}}}
        $
        is then calculated as follows:
        \begin{equation}
            \label{equ:modelling:phase field model:onsagers principle:second time derivative}
            \begin{split}
                \integral{\domain}{}{
                    \ginzburgLandauEnergyDens{\cortexPhaseField{}{}}{}
                    \diff{t}{}{
                        \pfCouplingEnergyDens{x,y}{
                            \pfSpeciesOne{}{},
                            \normal[\cortexPhaseField{}{}]{}
                        }
                    }
                }{\lebesgueM{3}}
                &=
                \integral{\domain}{}{
                    \ginzburgLandauEnergyDens{\cortexPhaseField{}{}}{}
                    \left(
                        \pDiff{\pfSpeciesOne{}{}}{}{}\pfCouplingEnergyDens{}{}
                        \pDiff{t}{}{}\pfSpeciesOne{}{}
                        +
                        \grad{\normal{}}{}{}
                        \pfCouplingEnergyDens{}{}
                        \cdot
                        \pDiff{t}{}{}\normal[\cortexPhaseField{}{}]{}
                    \right)
                }{\lebesgueM{3}}
                \\
                &\overset{(3)}{=}
                \integral{\domain}{}{
                    \pDiff{\pfSpeciesOne{}{}}{}{} \pfCouplingEnergyDens{}{}
                    \left(
                        \meanCurv{\cortexPhaseField{}{}}{}
                        \symVelocityNormal{\cortexPhaseField{}{}}
                        \pfSpeciesOne{}{}
                        -
                        \diver{}{}{\ginzburgLandauEnergyDens{\cortexPhaseFieldSym}{}
                          \pfSpeciesOne{}{}
                          \pfVelocity{}{}_\tau
                        }
                        +
                        R(\pfSpeciesOne{}{},\pfSpeciesTwo{}{},\membrPhaseField{}{},\cortexPhaseField{}{})
                    \right)
                }{\lebesgueM{3}}
                \\
                &+
                \integral{\domain}{}{
                    \ginzburgLandauEnergyDens{\cortexPhaseField{}{}}{}
                    \grad{\normal{}}{}{}
                    \pfCouplingEnergyDens{x,y}{}
                    \cdot
                    \frac{1}{\abs{\grad{}{}{}\cortexPhaseField{}{}}}
                    \orthProjMat{\normal[\cortexPhaseField{}{}]{}}{}
                    \grad{}{}{
                        \pDiff{t}{}{}\cortexPhaseField{}{}
                    }
                }{\lebesgueM{3}}
                \\
                &\overset{(4)}{=}
                \integral{\domain}{}{
                    \pDiff{\pfSpeciesOne{}{}}{}{} \pfCouplingEnergyDens{}{}
                    \left(
                        \meanCurv{\cortexPhaseField{}{}}{}
                        \symVelocityNormal{\cortexPhaseField{}{}}
                        \pfSpeciesOne{}{}
                        -
                        \diver{}{}{\ginzburgLandauEnergyDens{\cortexPhaseFieldSym}{}
                          \pfSpeciesOne{}{}
                          \pfVelocity{}{}_\tau
                        }
                        +
                        R(\pfSpeciesOne{}{},\pfSpeciesTwo{}{},\membrPhaseField{}{},\cortexPhaseField{}{})
                    \right)
                }{\lebesgueM{3}}
                \\
                &-
                \integral{\domain}{}{
                    \diver{}{}{  
                        \ginzburgLandauEnergyDens{\cortexPhaseField{}{}}{}
                        \transposed{
                            \grad{\normal{}}{}{}
                            \pfCouplingEnergyDens{x,y}{}
                        }
                        \frac{1}{\abs{\grad{}{}{}\cortexPhaseField{}{}}}
                        \orthProjMat{\normal[\cortexPhaseField{}{}]{}}{}
                    }
                    \pDiff{t}{}{}\cortexPhaseField{}{}
                }{\lebesgueM{3}},
            \end{split}
        \end{equation}
        \begin{itemize}
            \item[(3)] \eqref{equ:onsagers principle:%
            species one cont equ} and a small computation for the time derivative of the normal
            \item[(4)] Integrating by parts,  and using the homogeneous boundary conditions
            on the species density \eqref{equ:modelling:phase field:species one Dirichlet condition}
        \end{itemize}
        where we abbreviated
        $$
            \orthProjMat{\normal[\cortexPhaseField{}{}]{}}{} = 
            \identityMatrix[3] - \frac{
                \grad{}{}{}\cortexPhaseFieldSym \tensorProd \grad{}{}{}\cortexPhaseFieldSym
            }{\abs{\grad{}{}{}\cortexPhaseFieldSym}^2}
        $$
        and
        $$
            R(\pfSpeciesOne{}{},\pfSpeciesTwo{}{},\membrPhaseField{}{},\cortexPhaseField{}{}) 
            = 
            \diver{}{}{
                \ginzburgLandauEnergyDens{\cortexPhaseField{}{}}{}
                \speciesOneDiffusiv{}
                \grad{}{}{}\speciesOne{}{}
            }
            +
            \ginzburgLandauEnergyDens{\cortexPhaseField{}{}}{}
            \reactionOne{\speciesOne{}{},\speciesTwo{}{},\membrPhaseField{}{}}{}.
        $$

        Inserting \eqref{equ:onsagers principle:cortex cont equ}
        reveals further
        \begin{equation*}
            \begin{split}
                \integral{\domain}{}{
                    \ginzburgLandauEnergyDens{\cortexPhaseField{}{}}{}
                    \diff{t}{}{\pfCouplingEnergyDens{x,y}{\pfSpeciesOne{}{},\cortexPhaseField{}{}}}
                }{\lebesgueM{3}}
                &=
                \int_{\domain}
                    \pDiff{\pfSpeciesOne{}{}}{}{} \pfCouplingEnergyDens{}{}
                    \left(
                        \meanCurv{\cortexPhaseField{}{}}{}
                        \symVelocityNormal{\cortexPhaseField{}{}}
                        \pfSpeciesOne{}{}
                        +
                        R(\pfSpeciesOne{}{},\pfSpeciesTwo{}{},\membrPhaseField{}{},\cortexPhaseField{}{})
                    \right)
                \\
                    &-
                    \diver{}{}{
                        \ginzburgLandauEnergyDens{\cortexPhaseField{}{}}{}
                        \transposed{\grad{\normal{}}{}{}\pfCouplingEnergyDens{}{}}
                        \frac{1}{\abs{\grad{}{}{}\cortexPhaseField{}{}}}
                        \orthProjMat{\normal[\cortexPhaseField{}{}]{}}{}
                    }
                    (
                        -
                        \velocity{}{}\cdot\grad{}{}{}\cortexPhaseField{}{}
                        +
                        \diver{}{}{}\symCortexPhaseFieldFlux
                    )
                \;\mathrm{d}\lebesgueM{3}
            \end{split}
        \end{equation*}

        We see that we have two types of terms in the energy
        time derivative: those that account for dissipation caused by the energy fluxes
        and that are connected to entropy change in the system,
        and terms that represent mechanical work done on the system or
        represent change in the chemical energy.
        The source of the mechanical work are the fluid particles.
        To have consistency with Newton's actio-reactio principle,
        we shall therefore choose $ K $ equal
        to the forces corresponding to the mechanical work, i.e.,
        \begin{equation*}
            \begin{split}
                K(z)
                \cdot
                \symVelocity(z)
                &=
                \left(
                    \grad{}{L^2}{}
                    \helfrichPFEnergyFunct{\membrPhaseFieldSym}{}(z)
                    \grad{}{}{}\symMembrPhaseField(z)
                    +
                    \grad{}{L^2}{}
                    \helfrichPFEnergyFunct{\cortexPhaseFieldSym}{}(z)
                    \grad{}{}{}\symCortexPhaseField(z)
                \right)
                \cdot
                \symVelocity(z)
                \\
                &-
                \phaseFieldParam
                \abs{\grad{}{}{}\symMembrPhaseField(z)}^2
                \symVelocityNormalVec{\symMembrPhaseField}(z)
                \cdot
                \grad{z}{}{}\pfCouplingIntCortex(z)
                +
                \symVelocityNormal{\symMembrPhaseField}(z)
                \meanCurv{\symMembrPhaseField}{z}
                \pfCouplingIntCortex(z)
                \\
                &-
                \integral{\domain}{}{
                    \ginzburgLandauEnergyDens{\symMembrPhaseField}{}(y)
                    \left(
                    \phaseFieldParam
                    \abs{\grad{}{}{}\symCortexPhaseField}^2(z)
                    \symVelocityNormalVec{\symCortexPhaseField}(z)
                    \cdot
                    \grad{z}{}{
                        \pfCouplingEnergyDens{\cdot,y}{\speciesOne{}{},
                          \normal[\cortexPhaseField{}{}]{}}
                    }
                    -
                    \symVelocityNormal{\symCortexPhaseField}(z)
                    \meanCurv{\symCortexPhaseField}{z}
                    \pfCouplingEnergyDens{z,y}{\speciesOne{}{},
                      \normal[\cortexPhaseField{}{}]{}}
                    \right)
                }{\lebesgueM{3}(y)}
                \\
                &-
                \integral{\domain}{}{
                    \ginzburgLandauEnergyDens{\membrPhaseField{}{}}{}(y)
                    \pDiff{\pfSpeciesOne{}{}}{}{} \pfCouplingEnergyDens{}{}
                    (z,y,\pfSpeciesOne{}{},\normal[\cortexPhaseFieldSym]{})
                    \meanCurv{\cortexPhaseField{}{}}{z}
                    \symVelocityNormal{\cortexPhaseField{}{}}(z)
                    \pfSpeciesOne{\NOARG}{z}
                }{\lebesgueM{3}(y)}
                \\
                &-
                \integral{\domain}{}{
                    \ginzburgLandauEnergyDens{\membrPhaseField{}{}}{}(y)
                    \grad{z}{}{\pDiff{\pfSpeciesOne{}{}}{}{} \pfCouplingEnergyDens{}{}
                    (\cdot,y,\pfSpeciesOne{}{},\normal[\cortexPhaseFieldSym]{})
                    }
                    \ginzburgLandauEnergyDens{\cortexPhaseFieldSym}{}
                    \cdot
                    \symVelocityTangVec
                    \pfSpeciesOne{}{}
                }{\lebesgueM{3}(y)}
                \\
                &-
                \integral{\domain}{}{
                    \ginzburgLandauEnergyDens{\membrPhaseField{}{}}{}(y)
                    \diver{}{}{
                        \ginzburgLandauEnergyDens{\cortexPhaseField{}{}}{}(z)
                        \transposed{\grad{\normal{}}{}{}\pfCouplingEnergyDens{}{}}
                        \frac{1}{\abs{\grad{}{}{}\cortexPhaseField{}{}}}
                        \orthProjMat{\normal[\cortexPhaseField{}{}]{}}{}
                    }
                    \velocity{\NOARG}{z}\cdot\grad{z}{}{}\cortexPhaseField{}{}
                }{\lebesgueM{3}(y)}
            \end{split}
        \end{equation*}
        leading to
        \begin{equation*}
            \begin{split}
                K(z)
                &=
                \grad{}{L^2}{}
                \helfrichPFEnergyFunct{\membrPhaseFieldSym}{}(z)
                \grad{}{}{}\symMembrPhaseField(z)
                +
                \grad{}{L^2}{}
                \helfrichPFEnergyFunct{\cortexPhaseFieldSym}{}(z)
                \grad{}{}{}\symCortexPhaseField(z)
                \\
                &-
                \phaseFieldParam
                \abs{\grad{}{}{}\symMembrPhaseField(z)}^2
                \normal[\symMembrPhaseField]{z}
                \tensorProd
                \normal[\symMembrPhaseField]{z}
                \grad{z}{}{}\pfCouplingIntCortex(z)
                +
                \meanCurv{\symMembrPhaseField}{z}
                \pfCouplingIntCortex(z)
                \normal[\membrPhaseFieldSym]{z}
                \\
                &-
                \integral{\domain}{}{
                    \ginzburgLandauEnergyDens{\symMembrPhaseField}{}(y)
                    \left(
                    \phaseFieldParam
                    \abs{\grad{}{}{}\symCortexPhaseField}^2(z)
                    \normal[\cortexPhaseFieldSym]{z}
                    \tensorProd
                    \normal[\cortexPhaseFieldSym]{z}
                    \grad{z}{}{
                        \pfCouplingEnergyDens{\cdot,y}{\speciesOne{}{},
                          \normal[\cortexPhaseField{}{}]{}}
                    }
                    -
                    \meanCurv{\symCortexPhaseField}{z}
                    \pfCouplingEnergyDens{z,y}{
                        \speciesOne{}{},
                        \normal[\cortexPhaseField{}{}]{}
                    }
                    \normal[\cortexPhaseFieldSym]{z}
                    \right)
                }{\lebesgueM{3}(y)}
                \\
                &-
                \integral{\domain}{}{
                    \ginzburgLandauEnergyDens{\membrPhaseField{}{}}{}(y)
                    \pDiff{\pfSpeciesOne{}{}}{}{} \pfCouplingEnergyDens{}{}
                    (z,y,\pfSpeciesOne{}{},\normal[\cortexPhaseFieldSym]{})
                    \meanCurv{\cortexPhaseField{}{}}{z}
                    \pfSpeciesOne{\NOARG}{z}
                    \normal[\cortexPhaseFieldSym]{z}
                }{\lebesgueM{3}(y)}
                \\
                &-
                \integral{\domain}{}{
                    \ginzburgLandauEnergyDens{\membrPhaseField{}{}}{}(y)
                    \orthProjMat{\normal[\cortexPhaseFieldSym]{}}{z}
                    \grad{z}{}{\pDiff{\pfSpeciesOne{}{}}{}{} \pfCouplingEnergyDens{}{}
                    (\cdot,y,\pfSpeciesOne{}{},\normal[\cortexPhaseFieldSym]{})
                    }
                    \ginzburgLandauEnergyDens{\cortexPhaseFieldSym}{}
                    \pfSpeciesOne{}{}
                }{\lebesgueM{3}(y)}
                \\
                &-
                \integral{\domain}{}{
                    \ginzburgLandauEnergyDens{\membrPhaseField{}{}}{}(y)
                    \diver{}{}{
                        \ginzburgLandauEnergyDens{\cortexPhaseField{}{}}{}(z)
                        \transposed{\grad{\normal{}}{}{}\pfCouplingEnergyDens{}{}}
                        \frac{1}{\abs{\grad{}{}{}\cortexPhaseField{}{}}}
                        \orthProjMat{\normal[\cortexPhaseField{}{}]{}}{}
                    }
                    \grad{z}{}{}\cortexPhaseField{}{}
                }{\lebesgueM{3}(y)}.
            \end{split}
        \end{equation*}
        We note,
        \begin{equation}
            \label{equ:modelling:phase field model:L2 grad cortex phase field coupling energy} 
            \begin{split}
                \grad{\cortexPhaseField{}{}}{L^2}{}
                \pfCouplingEnergy{\membrPhaseFieldSym,\cortexPhaseFieldSym,\pfSpeciesOne{}{}}{}
                (z)
                \grad{}{}{}\cortexPhaseField{}{}(z)
                \cdot
                \pfVelocity{\NOARG}{z}
                =
                &-
                \integral{\domain}{}{
                    \ginzburgLandauEnergyDens{\symMembrPhaseField}{}(y)
                    \phaseFieldParam
                    \abs{\grad{}{}{}\symCortexPhaseField}^2(z)
                    \symVelocityNormalVec{\symCortexPhaseField}(z)
                    \cdot
                    \grad{z}{}{
                        \pfCouplingEnergyDens{\cdot,y}{\speciesOne{}{},\cortexPhaseField{}{}}
                    }
                }{\lebesgueM{3}(y)}
                \\
                &+
                \integral{\domain}{}{
                      \ginzburgLandauEnergyDens{\symMembrPhaseField}{}(y)
                    \symVelocityNormal{\symCortexPhaseField}(z)
                    \meanCurv{\symCortexPhaseField}{z}
                    \pfCouplingEnergyDens{z,y}{\speciesOne{}{},\cortexPhaseField{}{}}
                }{\lebesgueM{3}(y)}
                \\
                &-
                \integral{\domain}{}{
                    \ginzburgLandauEnergyDens{\membrPhaseField{}{}}{}(y)
                    \diver{}{}{
                        \ginzburgLandauEnergyDens{\cortexPhaseField{}{}}{}(z)
                        \transposed{\grad{\normal{}}{}{}\pfCouplingEnergyDens{}{}}
                        \frac{1}{\abs{\grad{}{}{}\cortexPhaseField{}{}}}
                        \orthProjMat{\normal[\cortexPhaseField{}{}]{}}{}
                    }
                    \velocity{\NOARG}{z}\cdot\grad{z}{}{}\cortexPhaseField{}{}
                }{\lebesgueM{3}(y)}
            \end{split}
        \end{equation}
        and
        \begin{equation}
            \label{equ:modelling:phase field model:L2 grad membr phase field coupling energy} 
            \begin{split}
                \grad{\membrPhaseField{}{}}{L^2}{}
                \pfCouplingEnergy{\membrPhaseFieldSym,\cortexPhaseFieldSym,\pfSpeciesOne{}{}}{}
                \grad{}{}{}\membrPhaseField{}{}
                \cdot
                \velocity{}{}
                =
                -
                \phaseFieldParam
                \abs{\grad{}{}{}\symMembrPhaseField}^2
                \symVelocityNormalVec{\symMembrPhaseField}
                \cdot
                \grad{z}{}{}\pfCouplingIntCortex
                +
                \symVelocityNormal{\symMembrPhaseField}
                \meanCurv{\symMembrPhaseField}{}
                \pfCouplingIntCortex
            \end{split}
        \end{equation}
        as was computed in 
        the Appendix of \cite{Werner2021}.
        So
        \begin{align*}
            K\cdot\pfVelocity{}{}
            &=
            \grad{\membrPhaseField{}{}}{L^2}{}
            \helfrichPFEnergyFunct{\membrPhaseField{}{}}{}
            \grad{}{}{}\membrPhaseField{}{}
            \cdot
            \pfVelocity{}{}
            +
            \grad{\cortexPhaseField{}{}}{L^2}{}
            \helfrichPFEnergyFunct{\cortexPhaseField{}{}}{}
            \grad{}{}{}\cortexPhaseField{}{}
            \cdot
            \pfVelocity{}{}
            +
            \grad{\membrPhaseField{}{},\cortexPhaseField{}{}}{L^2}{}
            \pfCouplingEnergy{}{}
            \begin{pmatrix}
                \grad{}{}{}\membrPhaseField{}{}
                \cdot
                \pfVelocity{}{}
                \\
                \grad{}{}{}\cortexPhaseField{}{}
                \cdot
                \pfVelocity{}{}                
            \end{pmatrix}
            \\
            &-
            \integral{\domain}{}{
                \ginzburgLandauEnergyDens{\membrPhaseField{}{}}{}(y)
                \pDiff{\pfSpeciesOne{}{}}{}{}
                \pfCouplingEnergyDens{}{}
                (\cdot,y,\pfSpeciesOne{}{},\normal[\cortexPhaseFieldSym]{})
                \meanCurv{\cortexPhaseField{}{}}{}
                \symVelocityNormal{\cortexPhaseField{}{}}
                \pfSpeciesOne{}{}
            }{\lebesgueM{3}(y)}
            \\
            &-
            \integral{\domain}{}{
                \ginzburgLandauEnergyDens{\membrPhaseField{}{}}{}(y)
                \grad{x}{}{\pDiff{\pfSpeciesOne{}{}}{}{} \pfCouplingEnergyDens{}{}
                (\cdot,y,\pfSpeciesOne{}{},\normal[\cortexPhaseFieldSym]{})}
                \ginzburgLandauEnergyDens{\cortexPhaseFieldSym}{}
                \cdot
                \symVelocityTangVec
                \pfSpeciesOne{}{}
            }{\lebesgueM{3}(y)}.
        \end{align*}

        Now we are in the position to calculate the stationarity condition of
        the optimisation problem
        \eqref{equ:modelling:phase field model:Onsager optimisation problem},
        which reads
        \begin{equation*}
            \gatDeriv{
                \left(
                    \symVelocityFlux,
                    \symMembrPhaseFieldFlux,
                    \symCortexPhaseFieldFlux
                \right)
            }{}{
                \dissipFunct{}{}
            }{\tau,j_{\symMembrPhaseField},j_{\symCortexPhaseField}}
            =
            -
            \gatDeriv{
                \left(
                    \symVelocityFlux,
                    \symMembrPhaseFieldFlux,
                    \symCortexPhaseFieldFlux
                \right)
            }{}{
                \diff{t}{}{}\freeEnergyFunct{}{}
            }{\tau,j_{\symMembrPhaseField},j_{\symCortexPhaseField}}
        \end{equation*}
        for all 
        $ 
            \tau \in \left[\sobolevHSet{1}{\domain}\right]^{(3,3)},
            j_{\symMembrPhaseField}, 
            j_{\symCortexPhaseField} \in \sobolevHDivSet[0]{\domain}
        $ (consistent with
        \eqref{equ:modelling:phase field:flux condition}).
        In detail, we have
        \begin{equation}
            \label{equ:modelling:phase fiel model:onsagers principle:flux computations}
            \begin{split}
                \integral{\domain}{}{
                    \frac{\symVelocityFlux}{\viscosity}
                    \frobProd
                    \tau
                    +
                    \frac{\symMembrPhaseFieldFlux}{\membrPhaseFieldMob{\symMembrPhaseField}}
                    \cdot
                    j_{\symMembrPhaseField}
                    +
                    \frac{\symCortexPhaseFieldFlux}{\cortexPhaseFieldMob{\symCortexPhaseField}}
                    \cdot
                    j_{\symCortexPhaseField}
                }{\lebesgueM{3}}
                &=
                \integral{\domain}{}{
                    \symVelocity
                    \cdot
                    \diver{}{}{}
                    \tau
                }{\lebesgueM{3}}
                +
                \grad{\membrPhaseField{}{}}{L^2}{}
                \helfrichPFEnergyFunct{\membrPhaseFieldSym}{}
                \diver{}{}{}j_{\symMembrPhaseField}
                +
                \grad{\cortexPhaseField{}{}}{L^2}{}
                \helfrichPFEnergyFunct{\cortexPhaseFieldSym}{}
                \diver{}{}{}j_{\symCortexPhaseField}
                \\
                &+
                \integral{\domain}{}{
                    \pfCouplingIntCortex
                    \grad{\membrPhaseField{}{}}{L^2}{}
                    \ginzburgLandauEnergyDens{}{}
                    \diver{}{}{}j_{\symMembrPhaseField}
                }{\lebesgueM{3}}
                \\
                &+
                \integral{\domain\times\domain}{}{
                    \ginzburgLandauEnergyDens{\symMembrPhaseField}{}
                    \pfCouplingEnergyDens{\cdot,\cdot,\speciesOne{}{}}{}
                    \grad{\cortexPhaseField{}{}}{L^2}{}
                    \ginzburgLandauEnergyDens{}{}
                    \diver{}{}{}j_{\symCortexPhaseField}
                }{\lebesgueM{3}\otimes\lebesgueM{3}}
                \\
                &-
                \integral{\domain\times\domain}{}{
                    \ginzburgLandauEnergyDens{\membrPhaseField{}{}}{}
                    \diver{}{}{
                        \ginzburgLandauEnergyDens{\cortexPhaseField{}{}}{}
                        \transposed{\grad{\normal{}}{}{}\pfCouplingEnergyDens{}{}}
                        \frac{1}{\abs{\grad{}{}{}\cortexPhaseField{}{}}}
                        \orthProjMat{\normal[\cortexPhaseField{}{}]{}}{}
                    }
                    \diver{}{}{}j_{\cortexPhaseField{}{}}
                }{\lebesgueM{3}\otimes\lebesgueM{3}}
            \end{split}
        \end{equation}
        Integration by parts and the fundamental lemma of variations lead us to
        \begin{equation}
            \label{equ:modelling:thermodynamical fluxes}
            \begin{split}
                \symVelocityFlux 
                &=
                -
                \viscosity
                \grad{}{}{}\symVelocity,
                \\
                \symMembrPhaseFieldFlux
                &=-
                \membrPhaseFieldMob{\symMembrPhaseField}
                \left(
                \grad{}{}{
                    \grad{\membrPhaseFieldSym}{L^2}{}
                    \helfrichPFEnergyFunct{\membrPhaseFieldSym}{}
                }
                +
                \grad{}{}{
                    \grad{\symMembrPhaseField}{L^2}{}
                    \pfCouplingEnergy{\membrPhaseFieldSym,\cortexPhaseFieldSym,\pfSpeciesOne{}{}}{}
                }
                \right),
                \\
                \symCortexPhaseFieldFlux
                &=-
                \cortexPhaseFieldMob{\symCortexPhaseField}
                \left(
                \grad{}{}{
                    \grad{\cortexPhaseFieldSym}{L^2}{}
                    \helfrichPFEnergyFunct{\cortexPhaseFieldSym}{}
                }
                +
                \grad{}{}{
                    \grad{\symCortexPhaseField}{L^2}{}
                    \pfCouplingEnergy{\membrPhaseFieldSym,\cortexPhaseFieldSym,\pfSpeciesOne{}{}}{}
                }
                \right).
            \end{split}
        \end{equation}
        \begin{remark}
            Note that 
            $ 
                \diver{}{}{}\symVelocityFlux 
                =
                \diver{}{}{}
                \cauchyStressTens{}{} 
            $ for constant 
            viscosity.
        \end{remark}
        The resulting full system reads
        \begin{subequations}
            \label{equ:modelling:phase field:resulting PDEs}
            \begin{equation}
                \label{equ:modelling:phase field resulting PDEs:momentum balance}
                \fluidMassDens(\pDiff{t}{}{}\symVelocity 
                +
                (\symVelocity\cdot
                \grad{}{}{})
                \symVelocity)
                -
                \diver{}{}{
                    \viscosity
                    \left(
                        \grad{}{}{}\pfVelocity{}{}
                        +
                        \transposed{\grad{}{}{}\pfVelocity{}{}}
                    \right)
                    -
                    \pfPressure{}{}
                }
                = K,
            \end{equation}
            \begin{equation}
                \label{equ:modelling:phase field resulting PDEs:incompressibility}
                \diver{}{}{}\velocity{}{} = 0,
            \end{equation}
            \begin{equation}
                \label{equ:modelling:phase field resulting PDEs:first interface evolution}
                \pDiff{t}{}{}\symMembrPhaseField 
                +
                \symVelocity
                \cdot
                \grad{}{}{}
                \symMembrPhaseField
                =
                \diver{}{}{
                    \membrPhaseFieldMob{\symMembrPhaseField}
                    \left(
                    \grad{}{}{
                        \grad{\membrPhaseFieldSym}{L^2}{}
                        \helfrichPFEnergyFunct{\membrPhaseField{}{}}{}
                    }
                    +
                    \grad{}{}{
                        \grad{\symMembrPhaseField}{L^2}{}
                        \pfCouplingEnergy{\membrPhaseFieldSym,\cortexPhaseFieldSym,
                        \pfSpeciesOne{}{}}{}
                    }
                    \right)
                },
            \end{equation}
            \begin{equation}
                \label{equ:modelling:phase field:resulting PDEs:second interface evolution}
                \pDiff{t}{}{}\symCortexPhaseField 
                +
                \symVelocity
                \cdot
                \grad{}{}{}
                \symCortexPhaseField
                =
                \diver{}{}{
                    \cortexPhaseFieldMob{\symCortexPhaseField}
                    \left(
                    \grad{}{}{
                        \grad{\cortexPhaseFieldSym}{L^2}{}
                        \helfrichPFEnergyFunct{\cortexPhaseField{}{}}{}
                    }
                    +
                    \grad{}{}{
                        \grad{\symCortexPhaseField}{L^2}{}
                        \pfCouplingEnergy{\membrPhaseFieldSym,\cortexPhaseFieldSym,
                        \pfSpeciesOne{}{}}{}
                    }
                    \right)
                },
            \end{equation}
            \vspace{-.5cm}\begin{align}
                \label{equ:modelling:phase field:resulting PDEs:reaction-diffusion of species one}
                \ginzburgLandauEnergyDens{\cortexPhaseField{}{}}{}
                \pDiff{t}{}{}\pfSpeciesOne{}{}
                -
                \symVelocityNormal{\cortexPhaseField{}{}}
                \meanCurv{\cortexPhaseField{}{}}{}
                \pfSpeciesOne{}{}
                -
                \diver{}{}{
                    \ginzburgLandauEnergyDens{\cortexPhaseField{}{}}{}
                    \speciesOneDiffusiv{}
                    \grad{}{}{}\speciesOne{}{}
                }
                +
                \diver{}{}{
                    \ginzburgLandauEnergyDens{\cortexPhaseField{}{}}{}
                    \pfVelocity{}{}_\tau
                    \speciesOne{}{}
                }
                &=\\\nonumber
                \ginzburgLandauEnergyDens{\cortexPhaseField{}{}}{}
                (\repairRate\pfSpeciesTwo{}{} - \discRate{\membrPhaseFieldSym}\pfSpeciesOne{}{})
                &,
                \\
                \label{equ:modelling:phase field:resulting PDEs:reaction-diffusion of species two}
                \ginzburgLandauEnergyDens{\cortexPhaseField{}{}}{}
                \pDiff{t}{}{}\pfSpeciesTwo{}{}
                -
                \symVelocityNormal{\cortexPhaseField{}{}}
                \meanCurv{\cortexPhaseField{}{}}{}
                \pfSpeciesTwo{}{}
                -
                \diver{}{}{
                    \ginzburgLandauEnergyDens{\cortexPhaseField{}{}}{}
                    \speciesTwoDiffusiv{}
                    \grad{}{}{}\speciesTwo{}{}
                }
                +
                \diver{}{}{
                    \ginzburgLandauEnergyDens{\cortexPhaseField{}{}}{}
                    \pfVelocity{}{}_\tau
                    \speciesTwo{}{}
                }
                &=\\\nonumber
                \ginzburgLandauEnergyDens{\cortexPhaseField{}{}}{}
                (-\repairRate\pfSpeciesTwo{}{} + \discRate{\membrPhaseFieldSym}\pfSpeciesOne{}{})
                &,
            \end{align}
        \end{subequations}
        with boundary conditions
        \begin{subequations}
            \label{equ:modelling:phase field:resulting PDEs:boundary conditions}
            \begin{align}
                \label{equ:modelling:phase field:resulting PDEs:boundary conditions:%
                velocity}
                \restrFun{\pfVelocity{}{}}{\boundary{\domain}} &= 0,
                \\
                \label{equ:modelling:phase field:resulting PDEs:boundary conditions:%
                phase field}
                \restrFun{\pDiff{\normal{}}{}{}\membrPhaseField{}{}}{\boundary{\domain}} = 
                \restrFun{\pDiff{\normal{}}{}{}\cortexPhaseField{}{}}{\boundary{\domain}}
                &= 0,
                \\
                \label{equ:modelling:phase field:resulting PDEs:boundary conditions:%
                fluxes}
                \restrFun{\symMembrPhaseFieldFlux}{\boundary{\domain}}\cdot\normal{} =
                \restrFun{\symCortexPhaseFieldFlux}{\boundary{\domain}}\cdot\normal{} &= 0,
                \\
                \label{equ:modelling:phase field:resulting PDEs:boundary conditions:%
                species}
                \restrFun{\pfSpeciesOne{}{}}{\boundary{\domain}} = 
                \restrFun{\pfSpeciesTwo{}{}}{\boundary{\domain}} &= 0.
            \end{align}
        \end{subequations}
        \begin{remark}
            Existence of a reduced version of this system, where $ \cortexPhaseFieldSym $
            is considered as a time-dependent parameter, has been shown in
            \cite[Theorem~7]{Werner2021}.
        \end{remark}
    \subsection{Sharp Interface Model}
        For a diffuse interface PDE system as derived above, one may ask 
        whether its solutions approach solutions of a sharp interface PDE system in the limit
        $ \phaseFieldParam\searrow 0 $. In \cite{Werner+2021} a formal asymptotic 
        analysis is provided to answer this question.
        The system we approach this way is the following:
        \begin{subequations}
            \label{equ:modelling:sharp interface:strong PDEs}
            \begin{align}
                \label{equ:modelling:sharp interface classical PDE model:momentum balance}
                \fluidMassDens(\pDiff{t}{}{}\velocity{}{} 
                +
                \left( \velocity{}{} \cdot \grad{}{}{} \right) \velocity{}{})
                - 
                \diver{}{}{} \cauchyStressTens{}{}
                &=
                0
                &\text{in}\;\domain\setminus(\interfOne{t}\cup\interfTwo{t}),
                \\
                \label{equ:modelling:sharp interface classical PDE model:mass cons}
                \diver{}{}{}\velocity{}{} &= 0
                &
                \text{in}\;\domain\setminus(\interfOne{t}\cup\interfTwo{t}),
                \\
                \label{equ:modelling:sharp interface classical PDE model:vel dirichlet cond}
                \velocity{t}{} &= 0
                &\text{on}\;\boundary{\domain},
                \\
                \label{equ:modelling:sharp interface classical PDE model:no jump if one}
                \jump{\velocity{}{}}_{\interfOne{t}}
                &=
                0
                &\text{on}\;\interfOne{t},
                \\
                \label{equ:modelling:sharp interface classical PDE model:no jump if two}
                \jump{\velocity{}{}}_{\interfTwo{t}}
                &=
                0
                &\text{on}\;\interfTwo{t},
                \\
                \label{equ:modelling:sharp interface:classical PDE model:%
                Neumann condition membrane}
                -\jump{ \cauchyStressTens{}{}\normal{}}
                &= 
                -\forceDensNS{\speciesOne{}{}}{}{}
                &
                \text{on}\;\interfOne{t},
                \\
                \label{equ:modelling:sharp interface:classical PDE model:%
                Neumann condition cortex}
                -\jump{ \cauchyStressTens{}{}\normal{}}
                &= 
                -\forceDensNSInterfTwo{\speciesOne{}{}}{}{}
                &
                \text{on}\;\interfTwo{t},
                \\
                \label{equ:modelling:sharp interface:classical PDE model:%
                transport equation membrane}
                \pDiff{t}{}{} \membrLvlSetFun{}{}
                +
                \velocity{}{}
                \cdot
                \grad{}{}{}\membrLvlSetFun{}{}
                &=
                0
                &\text{in}\;\domain,
                \\
                \label{equ:modelling:sharp interface:classical PDE model:%
                transport equation cortex}
                \pDiff{t}{}{} \cortexLvlSetFun{}{}
                +
                \velocity{}{}
                \cdot
                \grad{}{}{}\cortexLvlSetFun{}{}
                &=
                0
                &\text{in}\;\domain,
                \\
                \label{equ:modelling:sharp interface:classical PDE model:%
                species one evolution}
                \pDiff{t}{}{}\speciesOne{}{} 
                -
                \meanCurv{\interfTwo{t}}{}
                \velocity{}{}\cdot\normal[\interfTwo{t}]{}
                \speciesOne{}{}
                -
                \diver{\interfTwo{t}}{}{
                    \speciesOneDiffusiv{}
                    \grad{}{}{}\speciesOne{}{}
                }
                +
                \diver{\interfTwo{t}}{}{\speciesOne{}{}\symVelocityTangVec}
                &=
                &
                \text{on}\;\interfTwo{t},\\\nonumber
                \repairRate\inactiveLinkers{}{}
                -
                \discRate{\membrLvlSetFun{}{}}
                \activeLinkers{}{}
                &&
                \\
                \label{equ:modelling:sharp interface:classical PDE model:%
                species two evolution}
                \pDiff{t}{}{}\speciesTwo{}{} 
                -
                \meanCurv{\interfTwo{t}}{}
                \velocity{}{}\cdot\normal[\interfTwo{t}]{}
                \speciesTwo{}{}
                - 
                \diver{\interfTwo{t}}{}{
                    \speciesTwoDiffusiv{}
                    \grad{}{}{}\speciesTwo{}{}
                }
                +
                \diver{\interfTwo{t}}{}{\speciesTwo{}{}\symVelocityTangVec}
                &=
                &\text{on}\;\interfTwo{t},\\\nonumber
                -\repairRate\inactiveLinkers{}{}
                +
                \discRate{\membrLvlSetFun{}{}}
                \activeLinkers{}{}
        \end{align}
        \end{subequations}
        where
        \begin{align*}
            \forceDensNS{\speciesOne{}{}}{}{}                
            =
            -
            \grad{\membrLvlSetFun{}{}}{L^2}{}
            \helfrichPFEnergyFunct{}{}
            \grad{}{}{}\membrLvlSetFun{}{}
            +
            \left(\grad{y}{}{}\couplingIntCortex\cdot\normal[\interfOne{}]{}\right)
            \normal[\interfOne{}]{}
            -
            \meanCurv{\interfOne{}}{}
            \couplingIntCortex
            \normal[\interfOne{}]{},
        \end{align*}
        and
        \begin{align*}
            \forceDensNSInterfTwo{\speciesOne{}{}}{}{} 
            = 
            &
            -
            \grad{\cortexLvlSetFun{}{}}{L^2}{}
            \helfrichPFEnergyFunct{}{}
            \grad{}{}{}\cortexLvlSetFun{}{}
            +
            \left(\grad{x}{}{}\couplingIntMembr\cdot\normal[\interfTwo{}]{}\right)
            \normal[\interfTwo{}]{}
            -
            \meanCurv{\interfTwo{}}{}
            \couplingIntMembr
            \normal[\interfTwo{}]{}
            \\
            &+
            \pDiff{\speciesOne{}{}}{}{}
            \couplingIntMembr
            \meanCurv{\interfTwo{}}{}
            \speciesOne{}{}
            \normal[\interfTwo{}]{}
            +
            \grad{\interfTwo{}}{}{
            \pDiff{\speciesOne{}{}}{}{}
            \couplingIntMembr}
            \speciesOne{}{}
            \\
            &+
            \diver{\interfTwo{}}{}{
                \grad{\normal{}}{}{}\couplingIntMembr
            }
            \normal[\interfTwo{}]{}
            +
            \meanCurv{\interfTwo{}}{}
            \left(
            \grad{\normal{}}{}{}\couplingIntMembr
            \cdot
            \normal[\interfTwo{}]{}
            \right)
            \normal[\interfTwo{}]{}
        \end{align*}
        with
        $$
            \couplingIntCortex(y) = 
            \integral{\interfTwo{}}{}{
                \couplingEnergyDens{
                  x,y
                }{
                  \activeLinkers{}{},
                  \normal[\interfTwo{}]{}
                }
            }{\hausdorffM{2}(x)}
            \quad\text{and}\quad
            \couplingIntMembr(x) = 
            \integral{\interfOne{}}{}{
                \couplingEnergyDens{
                  x,y
                }{
                  \activeLinkers{}{},
                  \normal[\interfTwo{}]{}
                }
            }{\hausdorffM{2}(y)}.
        $$
        Note that $ \helfrichPFEnergyFunct{}{} $ is the Canham--Helfrich energy
        on surfaces and $ \membrLvlSetFun{}{} $, $ \cortexLvlSetFun{}{} $ are 
        level set functions for membrane and cortex, respectively.


\section{Numerical Experiments}
	We present numerical results in two dimensions for simulating bleb formation in cells
    according to the PDE model \eqref{equ:modelling:phase field:resulting PDEs},
    \eqref{equ:modelling:phase field:resulting PDEs:boundary conditions}.
    In our numerical experiments, we use \eqref{equ:ripping density model:evans}
    as a model for the ripping density together with a Gaussian distribution of 
    connection directions around the cortex normal \eqref{equ:angle dependent connectivity},
    \eqref{equ:Gaussian direction distribution}. The cytosol and the extracellular fluid 
    are assumed to be water at $20^\circ\mathrm{C}$.
    The biological literature offers quantitative results for most of the parameters involved;
    we have listed those on which our simulations are based in 
    Table~\ref{tab:numerical methods:applications:parameter choice}.
    \begin{table}
        \centering
        \begin{tabular}{l | c | c | l | l}
            Parameter & Symbol & Value & Unit & Reference 
            \\
            \midrule
            Fluid's viscosity & $ \viscosity $ & $ 1.006\cdot10^{-3} $ & $ \SIPascal\,\SISecond $ & 
            \cite[p.~1840]{Charras+2008}
            \\
            Fluid's density & $ \fluidMassDens $ & $ 998.2071 $ & $ \SIMass\,\SIMeter^{-3} $ & see text
            \\
            Temperature & $ \temperature $ & $ 293.15 $ & $ \SIKelvin $ & see text
            \\
            Char. energ. len. & $ \charEnergyLen $ & $ 0.1 $ & $ \mathrm{n}\SIMeter $ & \cite[p.~112]{Evans2001}
            \\
            Linker stiffness & $ \linkersSpringConst{} $ & $ 10^{-4} $ & $ \SINewton\SIMeter^{-1} $ 
            & \cite[Figure~2]{Yao+2016}
            \\
            Linker reconnection rate & $ \repairRate $ & $ 10^4 $ & $ \SISecond^{-1} $ &
            \cite[p.~1882]{Alert+2015}
            \\
            Attempt frequency & $ \koff $ & $ 10 $ & $ \SISecond^{-1} $ &
            \cite[p.~1882]{Alert+2015}
            \\
            Surface tension & $ \elComprMod $ & $ 5\cdot10^{-7} $ & $ \SIJoule\SIMeter^{-2} $ & 
            \cite[p.~177]{Safran+2005}
            \\
            Bending rigidity & $ \bendRig $ & $ 2\cdot10^{-20} $ & $ \SIJoule $ & 
            \cite[p.~176]{Safran+2005}
        \end{tabular}
        \caption{Typical parameters for cell blebbing.}
        \label{tab:numerical methods:applications:parameter choice}
    \end{table}  
    All presented simulations are carried out for a static cortex, i.e.,
    \eqref{equ:modelling:phase field:resulting PDEs:second interface evolution} is dropped
    together with the transport terms for the linker densities.

    \newcommand{\refLen}{x_r}
    \newcommand{\refTime}{t_r}
    \newcommand{\refMass}{m_r}
    \newcommand{\refVel}{\velocitySym_r}
    \newcommand{\refForce}{f_r}
    \newcommand{\refPress}{\pressure{}{}_r}
    \newcommand{\refCouplingEnergy}{C_r}
    \newcommand{\refMob}{M_r}
    With respect to a reference length of $ x_r = 10~\mu\SIMeter$ 
    (typical scale for cell diameters) and a reference
    time of $ \refTime = 30~\SISecond $ (time for bleb nucleation, cf.~\cite{Charras+2008})
    giving a reference
    velocity $ \refVel \approx 3.33\cdot10^{-7} $ and
    the parameters in Table~\ref{tab:numerical methods:applications:parameter choice},
    we non-dimensionalize. 
    The resulting system in weak formulation reads
    \begin{subequations}
        \begin{equation*}
            \begin{split}
                \mathrm{Re}
                \VLTProd{\domain}{3}{
                    \pDiff{t}{}{}\velocity{}{}
                    +
                    (\velocity{}{}\cdot\grad{}{}{})\velocity{}{}
                }{\testFunVelocity{}{}}
                +
                \MLTProd{\domain}{3}{
                    \grad{}{}{}\velocity{}{}
                }{
                    \grad{}{}{}\testFunVelocity{}
                }
                -
                \LTProd{\domain}{\pfPressure{}{}}{\diver{}{}{}\testFunVelocity{}}
                &=\\
                \nonumber
                \VLTProd{\domain}{3}{
                     \left(
                       	\frac{1}{\capillaryNum}
                        \genGlChemPotSym
                        +
                        \frac{1}{\relNumBend}
                        \genWillmChemPotSym
                        +
                        \frac{1}{\relNumCpl}
                        \grad{\membrPhaseFieldSym}{L^2}{}
                        \couplingEnergy{
                            \membrPhaseFieldSym,
                            \pfSpeciesOne{}{},
                            \cortexPhaseField{}{}
                        }{}
                    \right)
                    \grad{}{}{}\membrPhaseFieldSym
                    +
                    \extForceStokesSym
                }{\testFunVelocity{}},
                \\
                \LTProd{\domain}{\diver{}{}{}\pfVelocity{}{}}{\testFunPressure{}} &= 0,
            \end{split}
        \end{equation*}
        \begin{equation*}
            \begin{split}
            \dualProd[\topoDual{\sobolevHSet{1}{\domain}}]{
                \pDiff{t}{}{}\membrPhaseField{}{}
            }{
                \testFunMembrPhaseField{}
            }
            +
            \LTProd{\domain}{
                \velocity{}{}
                \cdot
                \grad{}{}{}
                \membrPhaseField{}{}
            }{
                \testFunMembrPhaseField{}
            }
            &=\\
            \nonumber
            -
            \VLTProd{\domain}{3}{
                \grad{}{}{
                  	\frac{1}{\pecletNum}
                    \genWillmChemPotSym
                    +
                    \frac{1}{\transpNumBend}
                    \genGlChemPotSym
                    +
                    \frac{1}{\transpNumCpl}
                    \grad{\membrPhaseField{}{}}{L^2}{}
                    \couplingEnergy{
                        \membrPhaseFieldSym,
                        \pfSpeciesOne{}{},
                        \cortexPhaseField{}{}
                    }{}
                }
            }{
                \grad{}{}{}
                \testFunMembrPhaseField{}
            }&,
            \end{split}
        \end{equation*}
        \vspace{-.5cm}\begin{align*}
            \LTProd{\domain}{
                \genGlChemPotSym
            }{
                \testFunGlChemPot{}
            }
            &=
            \cahnNum^2
            \VLTProd{\domain}{3}{
                \grad{}{}{}\membrLvlSetFunSym
            }{
                \grad{}{}{}\testFunGlChemPot{}
            }
            +
            \LTProd{\domain}{
                \doubleWellPot[\schemeSupscr{\prime}]{\membrLvlSetFunSym}
            }{
                \testFunGlChemPot{}
            },
            \\
            \LTProd{\domain}{
                \genWillmChemPotSym
            }{
                \testFunGenWillmChemPot{}
            }
            &=
            \cahnNum^2
            \VLTProd{\domain}{3}{
                \grad{}{}{
                    \genGlChemPotSym
                    +
                    \phaseFieldParam\pfSpontMeanCurv\left(\membrPhaseField{}{}^2-1\right)
                }
            }{
                \grad{}{}{}\testFunGenWillmChemPot{}
            }
            +\\\nonumber
            &\LTProd{\domain}{
                \left(
                    \genGlChemPotSym
                    +
                    \phaseFieldParam\pfSpontMeanCurv\left(\membrPhaseField{}{}^2-1\right)                         
                \right)
                \left(
                \doubleWellPot[\schemeSupscr{\prime\prime}]{\membrPhaseField{}{}}
                +
                2
                \phaseFieldParam
                \pfSpontMeanCurv
                \membrPhaseField{}{}
                \right)
            }{
                \testFunGenWillmChemPot{}
            },
        \end{align*}
        \vspace{-.5cm}\begin{align*}
            \mathrm{Pe}_a
            \dualProd[{
                    \sobolevWeightedHSet{
                        \ginzburgLandauEnergyDens{\cortexPhaseFieldSym_t}{}
                    }{-1}{\domain}
            }]{
                \pDiff{t}{}{}\pfSpeciesOne{}{}
            }{
                \testFunSpeciesOne{}
            }
            +
            \wVLTProd{\domain}{\ginzburgLandauEnergyDens{\cortexPhaseFieldSym_t}{}}{3}{
                \speciesOneDiffusiv{}
                \grad{}{}{}
                \pfSpeciesOne{}{}
            }{
                \grad{}{}{}
                \testFunSpeciesOne{}
            }
            &=
            \nonumber
            \wLTProd{\domain}{\ginzburgLandauEnergyDens{\cortexPhaseFieldSym_t}{}}{
              	\mathrm{Rec}_a\pfSpeciesTwo{}{}
                -
                \mathrm{Dis}_a\discRate{\membrPhaseFieldSym}\pfSpeciesOne{}{}
            }{
                \testFunSpeciesOne{}
            },
            \\
            \mathrm{Pe}_i
            \dualProd[{
                \sobolevWeightedHSet{
                    \ginzburgLandauEnergyDens{\cortexPhaseFieldSym_t}{}
                }{-1}{\domain}
            }]{
                \pDiff{t}{}{}\pfSpeciesTwo{}{}
            }{
                \testFunSpeciesTwo{}
            }
            +
            \wVLTProd{\domain}{\ginzburgLandauEnergyDens{\cortexPhaseFieldSym_t}{}}{3}{
                \speciesTwoDiffusiv{}
                \grad{}{}{}
                \pfSpeciesTwo{}{}
            }{
                \grad{}{}{}
                \testFunSpeciesTwo{}
            }
            &=
            \wLTProd{\domain}{\ginzburgLandauEnergyDens{\cortexPhaseFieldSym_t}{}}{
                -\mathrm{Rec}_i\pfSpeciesTwo{}{}
                +
                \mathrm{Dis}_i\discRate{\membrPhaseFieldSym}\pfSpeciesOne{}{}
            }{
                \testFunSpeciesTwo{}
            }
        \end{align*}
   \end{subequations}
    for all 
    $ 
        \testFunVelocity{} 
        \in 
        \sobolevHSet[0]{1}{\domain} 
    $,
    $ \testFunPressure{} \in \lebesgueSet{2}{\domain} $,
    $ 
        \testFunMembrPhaseField{},
        \testFunGlChemPot{},
        \testFunGenWillmChemPot{}
        \in 
        \sobolevHSet{1}{\domain}
    $,
    $ 
        \testFunSpeciesOne{}, \testFunSpeciesTwo{} 
        \in 
        \sobolevWeightedHSet{\ginzburgLandauEnergyDens{\cortexPhaseFieldSym_t}{}}{1}{\domain}.
    $

    We find 
    a small Reynolds number of about $ \mathrm{Re} \approx 3.3\cdot10^{-6} $, so 
    we neglect the inertia terms in the momentum balance 
    \eqref{equ:modelling:phase field resulting PDEs:momentum balance} and simulate 
    stationary Stokes flow instead of a Navier-Stokes system. 
    Other non-dimensional quantities are the capillary number $ \capillaryNum =
    \frac{\phaseFieldParam\viscosity}{\elComprMod\refTime} $ and 
    its pendants $ \relNumBend = \frac{\phaseFieldParam^3\viscosity}{\bendRig\refTime}, 
    \relNumCpl = \frac{\phaseFieldParam\viscosity}{\linkersSpringConst{}\refTime} $
    for the other energy components, 
    the Péclet number $ \pecletNum = \frac{\phaseFieldParam\refLen^2}{\elComprMod\refTime\refMob} 
    \approx 3.33 \cdot 10^{-3} $ and 
    its pendants $ \transpNumBend = \frac{\phaseFieldParam^3\refLen^2}{\bendRig\refTime\refMob} 
    \approx 2.08 \cdot 10^{-4} $, 
    $ \transpNumCpl =  \frac{\phaseFieldParam\refLen^2}{\linkersSpringConst{}\refTime\refMob} 
    \approx 1.67 \cdot 10^{-5}$ 
    for the other energy components, the Cahn number $ \cahnNum = \frac{\phaseFieldParam}{x_r} \approx 5 \cdot 10^{-3}$,
    Péclet numbers for the active and inactive linkers $ \mathrm{Pe}_a = 
    \frac{\refVel\refLen}{\speciesOneDiffusiv{}} $, 
    $\mathrm{Pe}_i = \frac{\refVel\refLen}{\speciesTwoDiffusiv{}} $
    as well as the relations of re- and disconnection to mass diffusion rates 
    $ \mathrm{Rec}_a = \frac{\repairRate\refLen^2}{\speciesOneDiffusiv{}}, 
    \mathrm{Rec}_i = \frac{\repairRate\refLen^2}{\speciesTwoDiffusiv{}}, 
    \mathrm{Dis}_a = \frac{\discRate{}\refLen^2}{\speciesOneDiffusiv{}}, 
    \mathrm{Dis}_i = \frac{\discRate{}\refLen^2}{\speciesTwoDiffusiv{}} $. 
    We shall point out that using parameters from a sharp interface setting 
    in a diffuse approach requires rescaling of the capillary number by $ \cahnNum $
    as has been mentioned in \cite{Ceniceros+2010}; this is also true for 
    $ \relNumCpl $ and a rescaling of $ \relNumBend$ is done by the factor $ \cahnNum^3 $.
    This leads to
    $ \capillaryNum \approx 6.71 \cdot 10^{-4} $, $ \relNumBend \approx 1.68 $, and
    $ \relNumCpl \approx 3.35\cdot10^{-6} $.
    
	\subsection{Scheme and Implementation}
    	For spatial discretization of the Stokes subsystem, we use Taylor--Hood 
        $ \mathbb{P}_2 $--$ \mathbb{P}_1 $ elements. The phase field $ \membrPhaseFieldSym $ 
        and the chemical potentials $ \genGlChemPotSym, \genWillmChemPotSym $, as well 
        as the linker densities $ \pfSpeciesOne{}{}, \pfSpeciesTwo{}{} $
        are approximated with $H^1$-conformal $ \mathbb{P}_1 $ elements.
    	
        The time is discretized semi-implicitly by a first order splitting scheme. To provide for discrete energy 
        stability, a secant method such as the one presented in \cite{Guillen-Gonzalez+2018} is employed
        using
        $$
            G_{\text{sec}}(\fdMembrPhaseFieldSym{n+1},\fdMembrPhaseFieldSym{n})
            \colonequals
            \frac{
                G(\fdMembrPhaseFieldSym{n+1})
                -
                G(\fdMembrPhaseFieldSym{n})                        
            }{
                \fdMembrPhaseFieldSym{n+1}-\fdMembrPhaseFieldSym{n}
            }
            \quad\text{for}\quad
        	G(\membrPhaseFieldSym)
            =
            \frac{1}{\phaseFieldParam}
            \doubleWellPot[\schemeSupscr{\prime}]{\membrPhaseFieldSym}
            +
            \pfSpontMeanCurv
            \left(\membrPhaseFieldSym^2-1\right).
        $$

        All together, the space and time-discrete scheme is (suppressing the non-dimensional 
        constants for a moment)
        \begin{subequations}
            \label{equ:numerical methods:numerical schemes:space-time discrete PDE}
            \begin{align}
                \label{equ:numerical methods:numerical schemes:space-time discrete PDE:%
                momentum balance}
                \viscosity
                \VLTProd{\domain_\meshSize}{(3,3)}{
                    \grad{}{}{}\fdVelocitySym{n+1}
                }{
                    \grad{}{}{}\testFunVelocity{}
                }
                -
                \LTProd{\domain_\meshSize}{\fdPressureSym{n+1}}{\diver{}{}{}\testFunVelocity{}}
                &=
                \\ 
                \nonumber
                \VLTProd{\domain_\meshSize}{3}{
                     \left(
                        \fdGenGlChemPotSym{n+1}
                        +
                        \fdGenWillmChemPotSym{n+1}
                        +
                        \grad{\membrPhaseFieldSym}{L^2}{}
                        \couplingEnergy{
                            \fdMembrPhaseFieldSym{n},
                            \fdSpeciesOneSym{n},
                            \cortexPhaseField{}{}
                        }{}
                    \right)
                    \grad{}{}{}\fdMembrPhaseFieldSym{n+1}
                    +
                    \extForceStokesSym^{n+1}
                }{\testFunVelocity{}},
                \\
                \label{equ:numerical methods:numerical schemes:space-time discrete PDE:%
                incomp}
                \LTProd{\domain_\meshSize}{\diver{}{}{}\fdVelocitySym{n+1}}{\testFunPressure{}}
                &= 
                0,
            \end{align}
            \begin{equation}
                \label{equ:numerical methods:numerical schemes:space-time discrete PDE:%
                phase field evol}
                \begin{split}
                \LTProd{\domain_\meshSize}{
                    \frac{ \fdMembrPhaseFieldSym{n+1} - \fdMembrPhaseFieldSym{n+1} }{\tau}
                }{
                    \testFunMembrPhaseField{}
                }
                +
                \LTProd{\domain_\meshSize}{
                    \fdVelocitySym{n+1} \cdot \grad{}{}{}\fdMembrPhaseFieldSym{n+1}
                }{
                    \testFunMembrPhaseField{}
                }
                =
                \\
                -
                \VLTProd{\domain_\meshSize}{3}{
                    \membrPhaseFieldMob{\fdMembrPhaseFieldSym{n+1}}
                    \grad{}{}{
                        \fdGenGlChemPotSym{n+1}{}
                        +
                        \fdGenWillmChemPotSym{n+1}{}
                        +
                        \grad{\membrPhaseFieldSym}{L^2}{}
                        \couplingEnergy{
                            \fdMembrPhaseFieldSym{n},
                            \cortexPhaseFieldSym,
                            \fdSpeciesOneSym{n}
                        }{}
                    }
                }{
                    \grad{}{}{}\testFunMembrPhaseField{}
                },
                \end{split}
            \end{equation}
            \vspace{-.5cm}\begin{align}
                \label{equ:numerical methods:numerical schemes:space-time discrete PDE:%
                gen gl chem pot}
                \LTProd{\domain_\meshSize}{
                    \fdGenGlChemPotSym{n+1}
                }{
                    \testFunGlChemPot{}
                }
                &=
                \VLTProd{\domain_\meshSize}{3}{
                    \phaseFieldParam
                    \grad{}{}{}\fdMembrPhaseFieldSym{n+1}
                }{
                    \grad{}{}{}\testFunGlChemPot{}
                }
                +
                \LTProd{\domain_\meshSize}{
                    \frac{1}{\phaseFieldParam}
                    \doubleWellPot[\schemeSupscr{\prime}]{\fdMembrPhaseFieldSym{n+1}}
                }{
                    \testFunGlChemPot{}
                },
                \\
                \label{equ:numerical methods:numerical schemes:space-time discrete PDE:%
                gen willm chem pot}
                \LTProd{\domain_\meshSize}{
                    \fdGenWillmChemPotSym{n+1}
                }{
                    \testFunGenWillmChemPot{}
                }
                &=
                \LTProd{\domain_\meshSize}{
                    \grad{}{}{
                        \discrGlGenChemPot{\fdGenGlChemPotSym{n+1}}{\fdMembrPhaseFieldSym{n+1}}
                    }
                }{
                    \grad{}{}{}\testFunGenWillmChemPot{}
                }
                \\\nonumber
                &+
                \frac{1}{\phaseFieldParam}
                \LTProd{\domain_\meshSize}{
                  \discrGlGenChemPot{
                    \fdGenGlChemPotSym{n+1}
                  }{
                    \fdMembrPhaseFieldSym{n+1}
                  }
                  G_{\text{sec}}(\fdMembrPhaseFieldSym{n+1},\fdMembrPhaseFieldSym{n})
                }{
                  \testFunGenWillmChemPot{}                      	
                },
            \end{align}
            \vspace{-.5cm}\begin{align}
                \label{equ:numerical methods:numerical schemes:space-time discrete PDE:%
                active species}
                \wLTProd{\domain_\meshSize}{
                    \ginzburgLandauEnergyDens{\cortexPhaseField{}{}}{}
                }{
                    \frac{\fdSpeciesOneSym{n+1} - \fdSpeciesOneSym{n}}{\timeStepSz}
                }{
                    \testFunSpeciesOne{}
                }
                +
                \wVLTProd{\domain_\meshSize}{
                    \ginzburgLandauEnergyDens{\cortexPhaseField{}{}}{}
                }{3}{
                    \speciesOneDiffusiv{}
                    \grad{}{}{}\fdSpeciesOneSym{n+1}
                }{
                    \grad{}{}{}\testFunSpeciesOne{}
                }
                &=\\\nonumber
                \wLTProd{\domain_\meshSize}{
                    \ginzburgLandauEnergyDens{\cortexPhaseField{}{}}{}
                }{
                  	\repairRate
                    \fdSpeciesTwoSym{n}
                    -
                    \discRate{\fdMembrPhaseFieldSym{n}}
                    \fdSpeciesOneSym{n}
                }{\testFunSpeciesOne{}}&,
                \\
                \label{equ:numerical methods:numerical schemes:space-time discrete PDE:%
                inactive species}
                \wLTProd{\domain_\meshSize}{
                    \ginzburgLandauEnergyDens{\cortexPhaseField{}{}}{}
                }{
                    \frac{\fdSpeciesTwoSym{n+1} - \fdSpeciesTwoSym{n}}{\timeStepSz}
                }{
                    \testFunSpeciesTwo{}
                }
                +
                \wVLTProd{\domain_\meshSize}{
                    \ginzburgLandauEnergyDens{\cortexPhaseField{}{}}{}
                }{3}{
                    \speciesTwoDiffusiv{}
                    \grad{}{}{}\fdSpeciesTwoSym{n+1}
                }{
                    \grad{}{}{}\testFunSpeciesTwo{}
                }
                &=\\\nonumber
                \wLTProd{\domain_\meshSize}{
                    \ginzburgLandauEnergyDens{\cortexPhaseField{}{}}{}
                }{
                    -\repairRate
                    \fdSpeciesTwoSym{n}
                    +
                    \discRate{\fdMembrPhaseFieldSym{n}}
                    \fdSpeciesOneSym{n}
                }{\testFunSpeciesTwo{}}&.
            \end{align}
        \end{subequations}
        
        This scheme is implemented using the FEM package NGSolve \cite{Schoeberl2014}.
        The overall algorithm also incorporates an inexact Newton method where the
        linearized system is solved iteratively using a combination of 
        a BDDC preconditioner and a GMRES method.

        A major difficulty regarding computational cost arises with 
        the force terms that result from the coupling energy 
        $ \pfCouplingEnergy{\membrPhaseFieldSym,\cortexPhaseFieldSym,\pfSpeciesOne{}{}}{} $
        (cf. \eqref{equ:numerical methods:numerical schemes:space-time discrete PDE:%
        momentum balance}, \eqref{equ:numerical methods:numerical schemes:space-time discrete PDE:%
        incomp}).
        This can be seen by computing the variation of this functional with respect 
        to $ \membrPhaseFieldSym $:
        \begin{align*}
            \grad{\membrPhaseFieldSym}{L^2}{}
            \pfCouplingEnergy{
              	\membrPhaseFieldSym,
                \cortexPhaseFieldSym,
                \pfSpeciesOne{}{}
            }{}(y)
            =
            &\frac{
            \genGlChemPot{\membrPhaseFieldSym}{}(y)}{2}
            \int_{\domain} 
                \ginzburgLandauEnergyDens{\cortexPhaseFieldSym}{}(x)
                \abs{x-y}^2
                \speciesOne{\NOARG}{x}
                \cdot\couplingProb{
                    \frac{(x-y)\cdot\normal[\interfTwo{}]{x}}{\abs{x-y}}
                }{}
            \mathrm{d}\lebesgueM{3}(x),
        \end{align*}
        where we abbreviate the $ L^2 $-gradient of the Ginzburg--Landau energy
        by $ \genGlChemPot{\membrPhaseFieldSym}{}
        = -\phaseFieldParam\laplacian{}{}{}\membrPhaseFieldSym + \frac{1}{\phaseFieldParam}
        \doubleWellPot{\membrPhaseFieldSym} $. These forces require the evaluation of 
        an integral over $ \domain $ in every point $ y \in \domain $, so the assembly
        routine runs in $ O(N^2)$, where $ N $ is the number of quadrature points in the mesh.
        To overcome issues of long simulation time, we precompute this term at the beginning
        of each time step and store the result on every quadrature point $ y $ in a cache,
        so that during the Newton iterations, we only execute look-up operations.
        The precomputation itself is conducted on a computation cluster combining MPI
        and multithreading.
    \subsection{Results}
    	We discuss two typical scenarios for cell blebbing
        both starting with the membrane resting on 
        the sphere $ \sphere{0.1} $ of radius $ 0.1 $, which is the cell cortex.
        In the first scenario, a directed force density 
        of magnitude $ \extForceStokesSc = 100 \, \SIPascal\, \SIMeter^{-1} $ 
        (see \cite{Charras+2008}) is applied
        that pushes out the membrane and causes destruction of the
        linkers. We neglect the forces of the linkers on the membrane here since 
        we are interested in the linker disconnection and final bleb height and shape.
        (Linker forces only account for an initial force barrier that has to be overcome;
        studies to find this barrier in the form of the so-called ``critical pressure''
        are presented in \cite{Werner+2020} for a related sharp interface model.)

        In the second scenario, we consider the case 
        of a homogeneous intracellular force density and a cortex that is destroyed at a 
        particular site. At this site, the linker densities are zero, and we 
        expect a bleb to develop there.
        In this scenario, we include the linker forces to show that they hold back the 
        membrane wherever they are present and a protrusion can only develop in their absence.
        \paragraph{Directed Force Density}
        	It has been hypothesized by \cite{Goudarzi+2017} that blebbing requires a folded 
            membrane, so-called invaginations, so there is enough material to be pushed out 
            and the surface tension does not totally prohibit the protrusion. To avoid modelling 
            a folded membrane, we account for this by rescaling the surface 
            tension $ \elComprMod $. A parameter study shows that a cell radius to 
            bleb height ratio that has been reported in the biological literature 
            (cf. \cite{Charras+2008}, \cite{Lee+2014})
            is reached at about $ \elComprMod = 5\cdot10^{-11} $, which conclusively
            is the surface tension that we employ in the following. In Figure~\ref{fig:directed %
            force}, the evolution of the phase field $ \fdMembrPhaseFieldSym{n} $ is shown
            at different (non-dimensional) time points. The force density applied is 
            \begin{equation}
                \label{equ:force density}
                \extForceStokesSym(x) = \extForceStokesSc
                \exp\left( -\measuredangle\left(x-m,(0,1)\right)^2 \mathbin{/} (20\cahnNum) \right) \cdot
                \exp\left(-\dist{m+\sphere{0.1}}{x}^2\mathbin{/} (20\cahnNum) \right),
            \end{equation}
            which can be thought of as a Gaussian in the midpoint $ m = (0.5,0.5)^T $ 
            that is concentrated around the ``northern'' direction $ (0,1)^T $.
            \begin{figure}
                    \includegraphics[scale=.28]{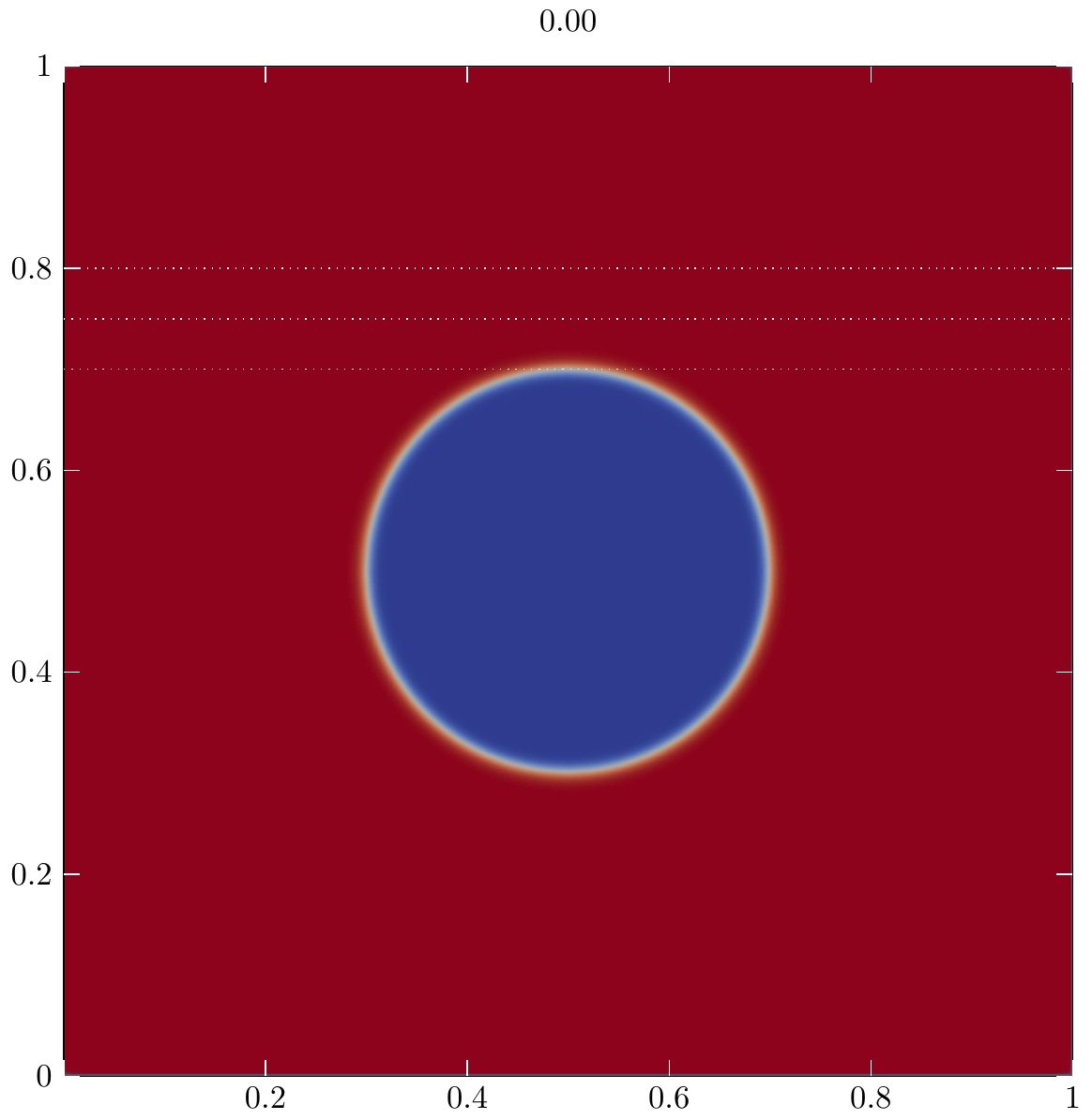}
                    \includegraphics[scale=.28]{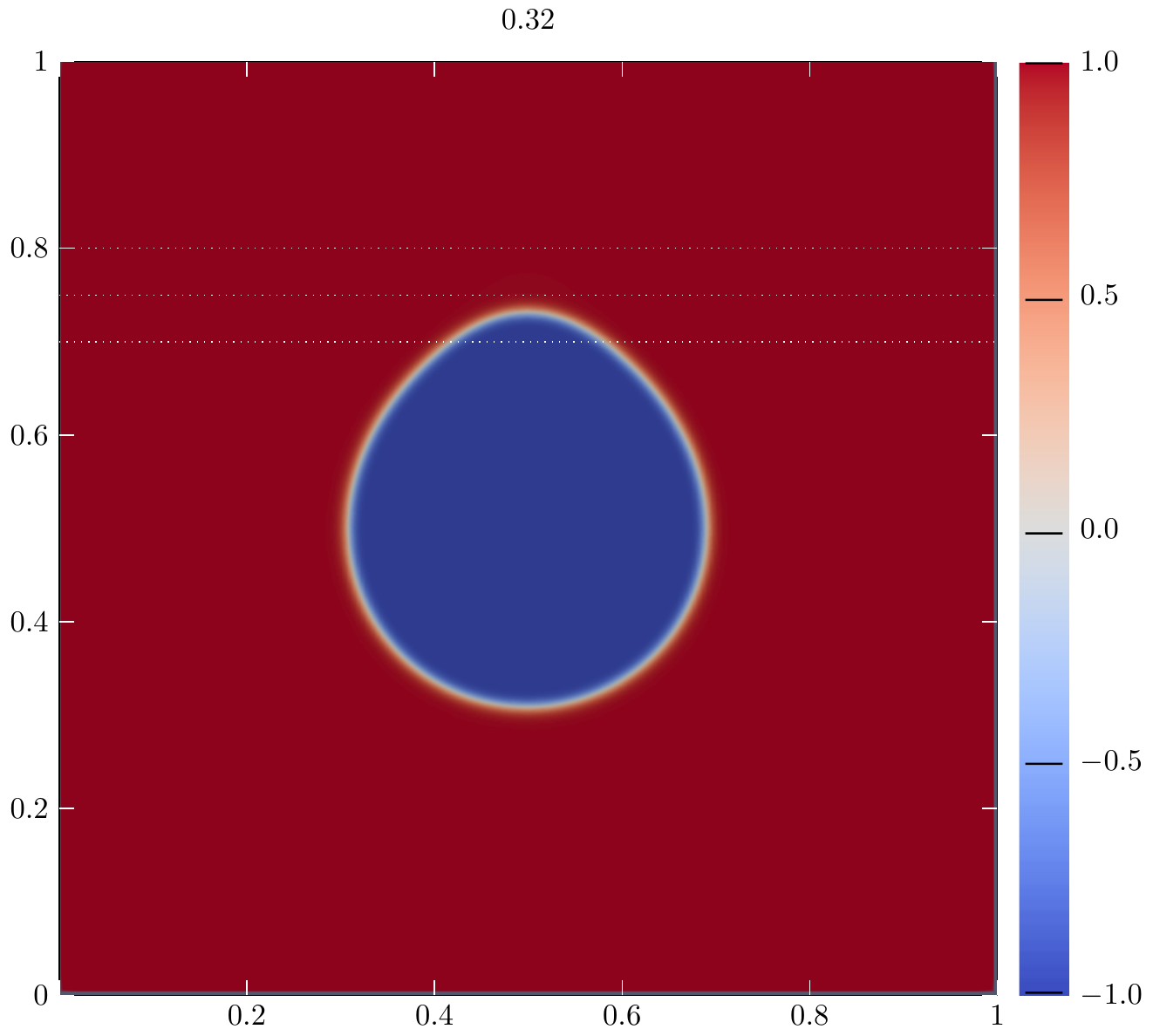}
                    \includegraphics[scale=.28]{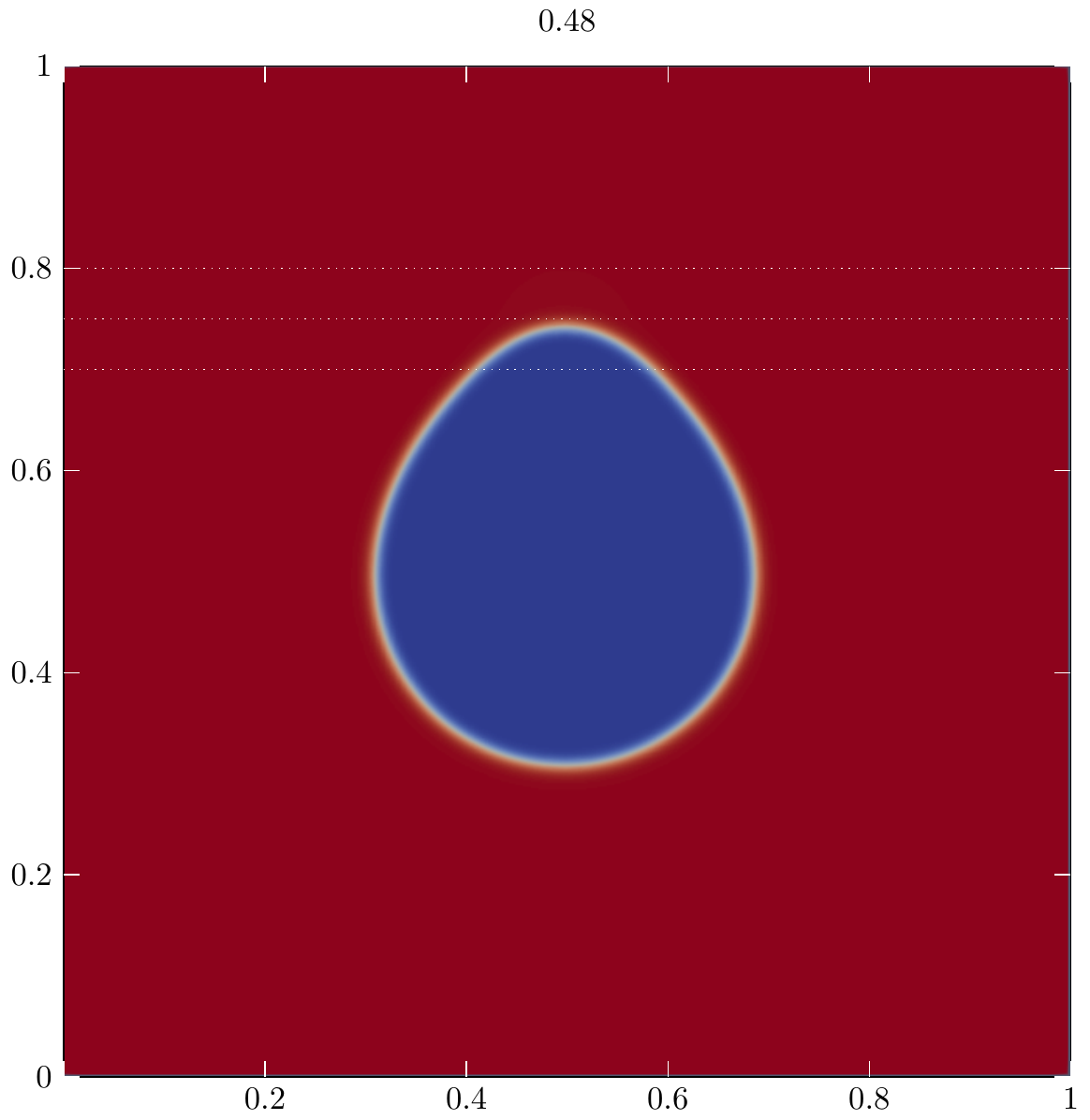}
               \includegraphics[scale=.28]{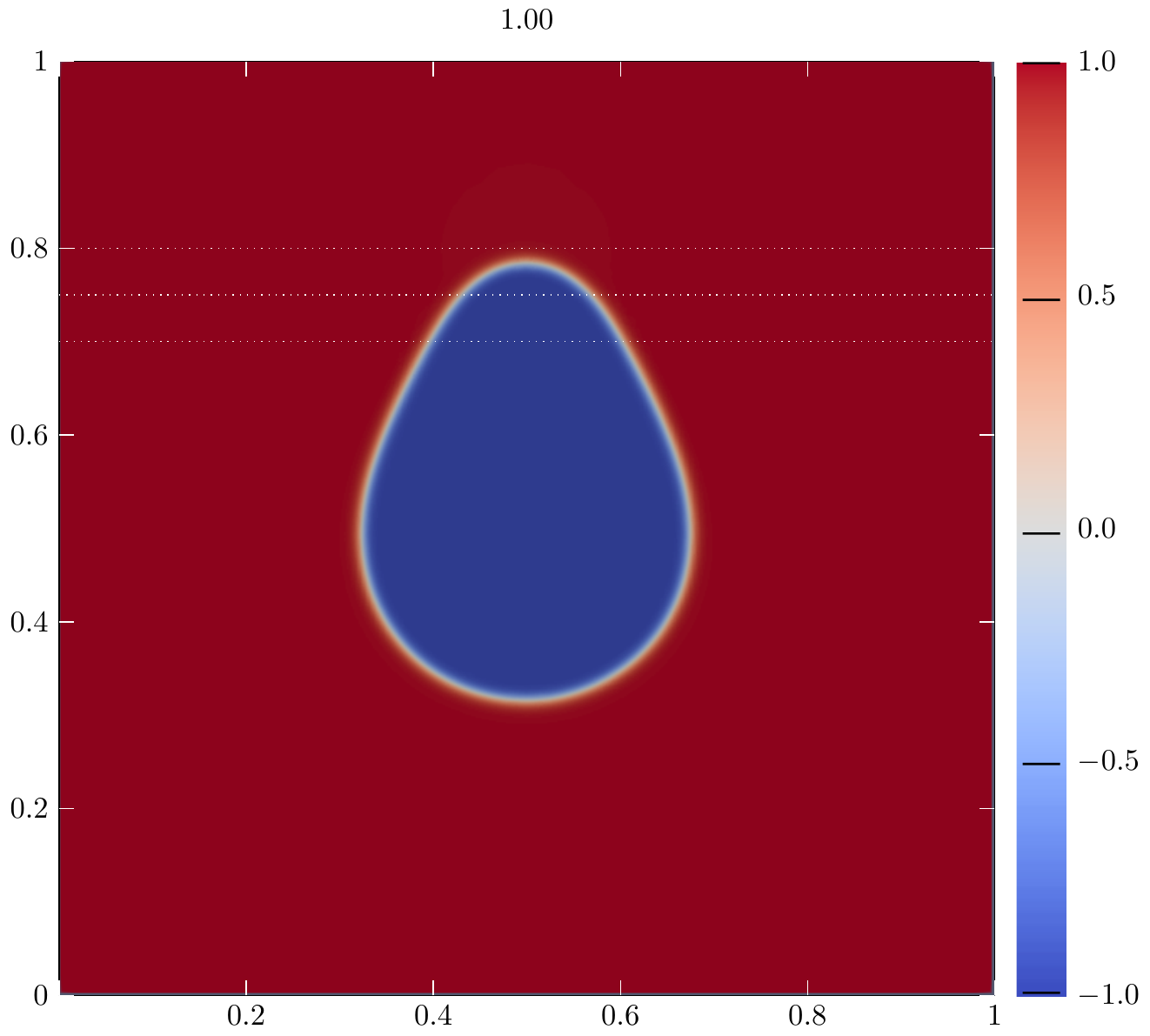}
                \caption{Contour plot of $ \fdMembrPhaseFieldSym{n} $ showing the
                nucleation and expansion of a bleb over non-dimensional time 
                with surface tension $ \elComprMod = 5 \cdot 10^{-11} \SIJoule\,\SIMeter^{-2} $.}
                \label{fig:directed force}
            \end{figure}
            We consider a reduced species evolution law
            \eqref{equ:numerical methods:numerical schemes:space-time discrete PDE:%
            active species},
            \eqref{equ:numerical methods:numerical schemes:space-time discrete PDE:%
            inactive species}:
            the density $ \fdSpeciesTwoSym{n} $ is exchanged with $ \fdSpeciesTwoSym{n} = \totalSpecDens
            - \fdSpeciesOneSym{n} $ for a total density of $ \totalSpecDens = 2\cdot10^{19}\,\SIMeter^{-3} $
            (which is the total density estimated in
            \cite[p.~1881]{Alert+2015} and rescaled to three dimensions
            by dividing with $ x_r $), and 
            the linkers' diffusivity is set to zero as well as the regeneration rate.
            In Figure~\ref{fig:linker disc}, we see how the density of active linkers 
            decreases while the bleb expands. 
            \begin{figure}
                \begin{subfigure}{.9\textwidth}
                    \includegraphics[scale=0.35]{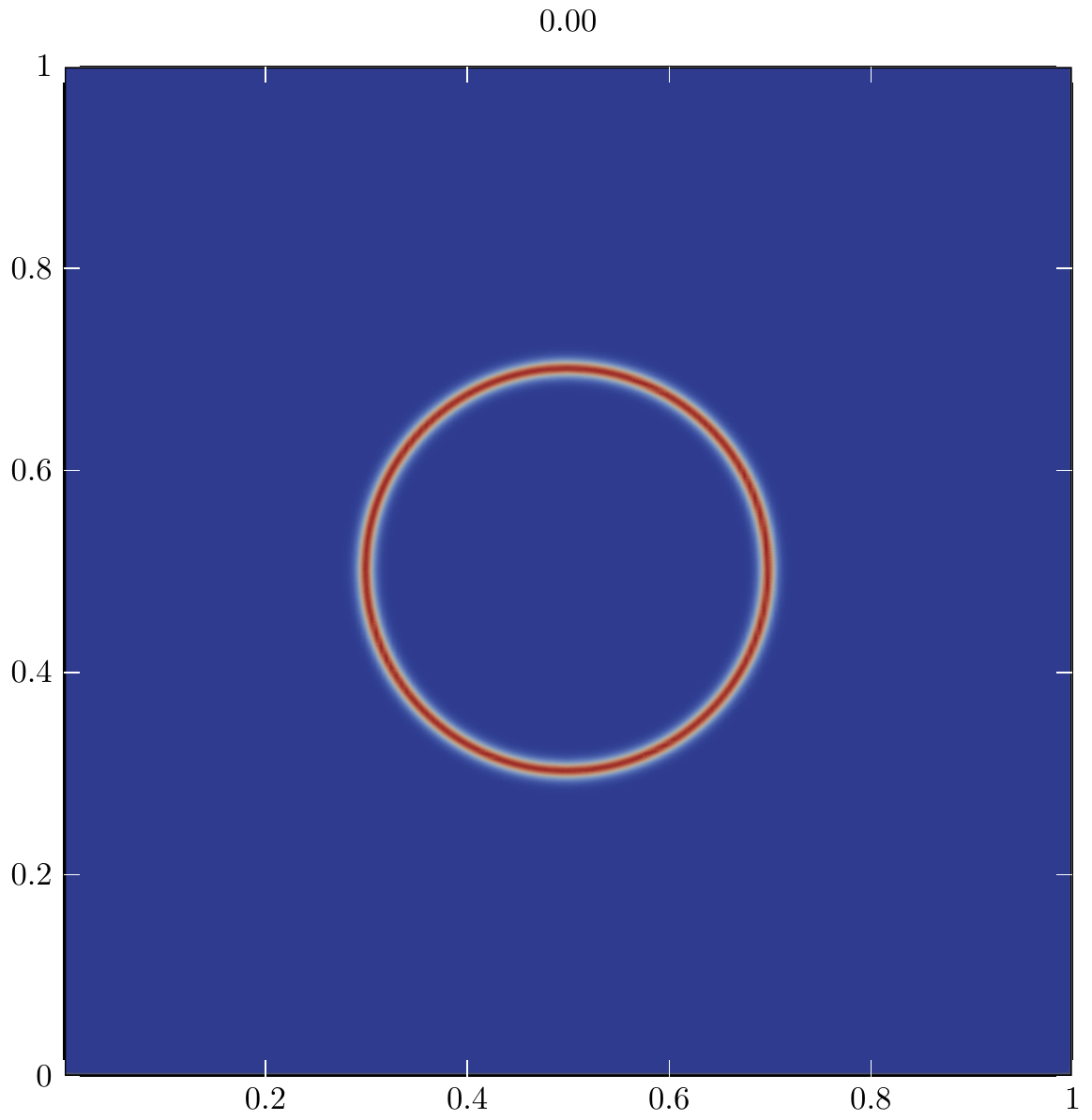}
                    \hfill
                    \includegraphics[scale=.35]{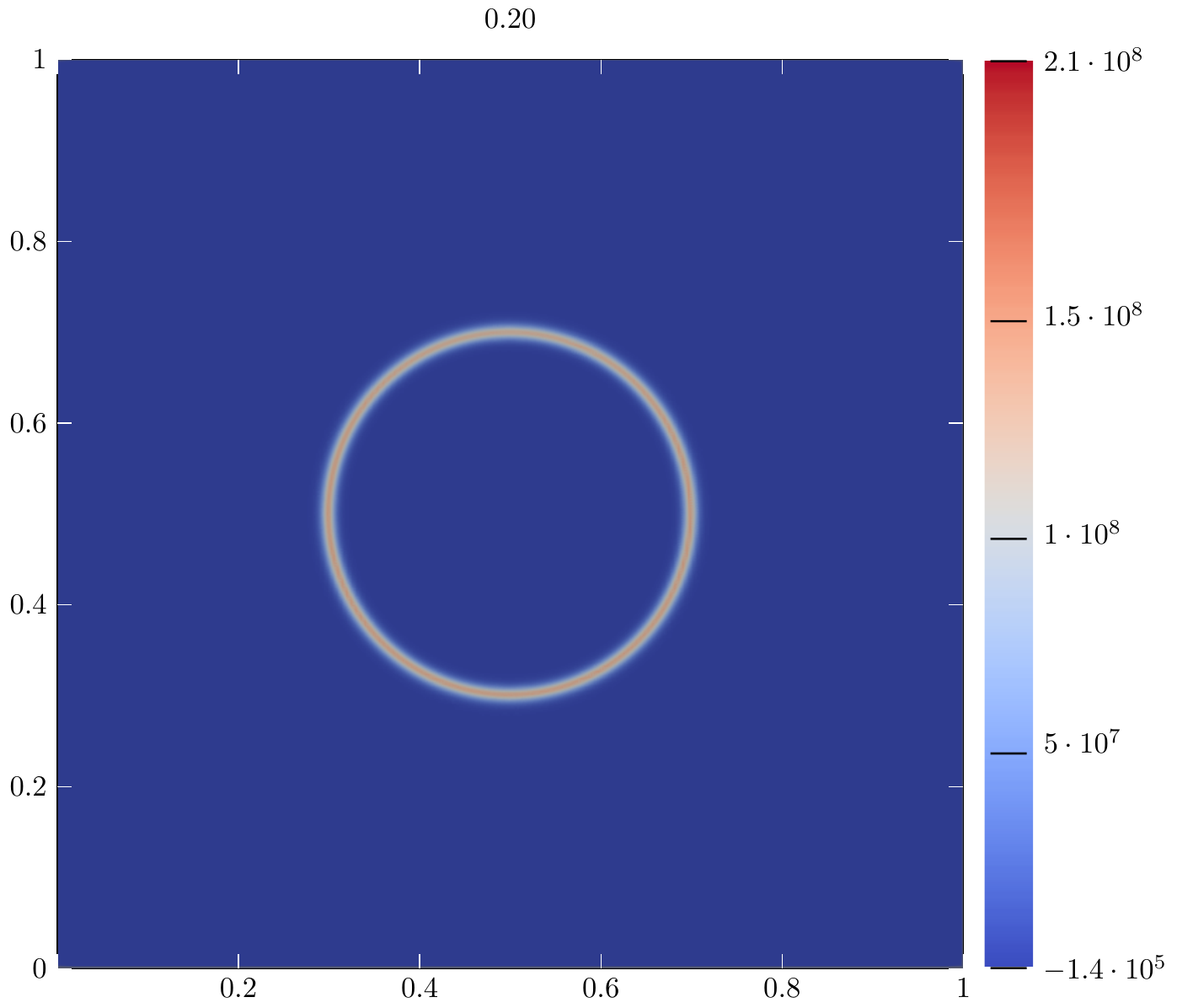}
                    \caption{Linker densities
                    with $ \elComprMod = 5\cdot10^{-11}~\SIJoule\,\SIMeter^{-2}$,
                    force $ \extForceStokesSc = 100~\SIPascal\,\SIMeter^{-1} $,
                    total linker density $ \totalSpecDens = 2.1\cdot10^{9}~\SIMeter^{-3} $.}
                    \label{fig:linker disc}
                \end{subfigure}
                \begin{subfigure}{.9\linewidth}
                    \includegraphics[scale=.35]{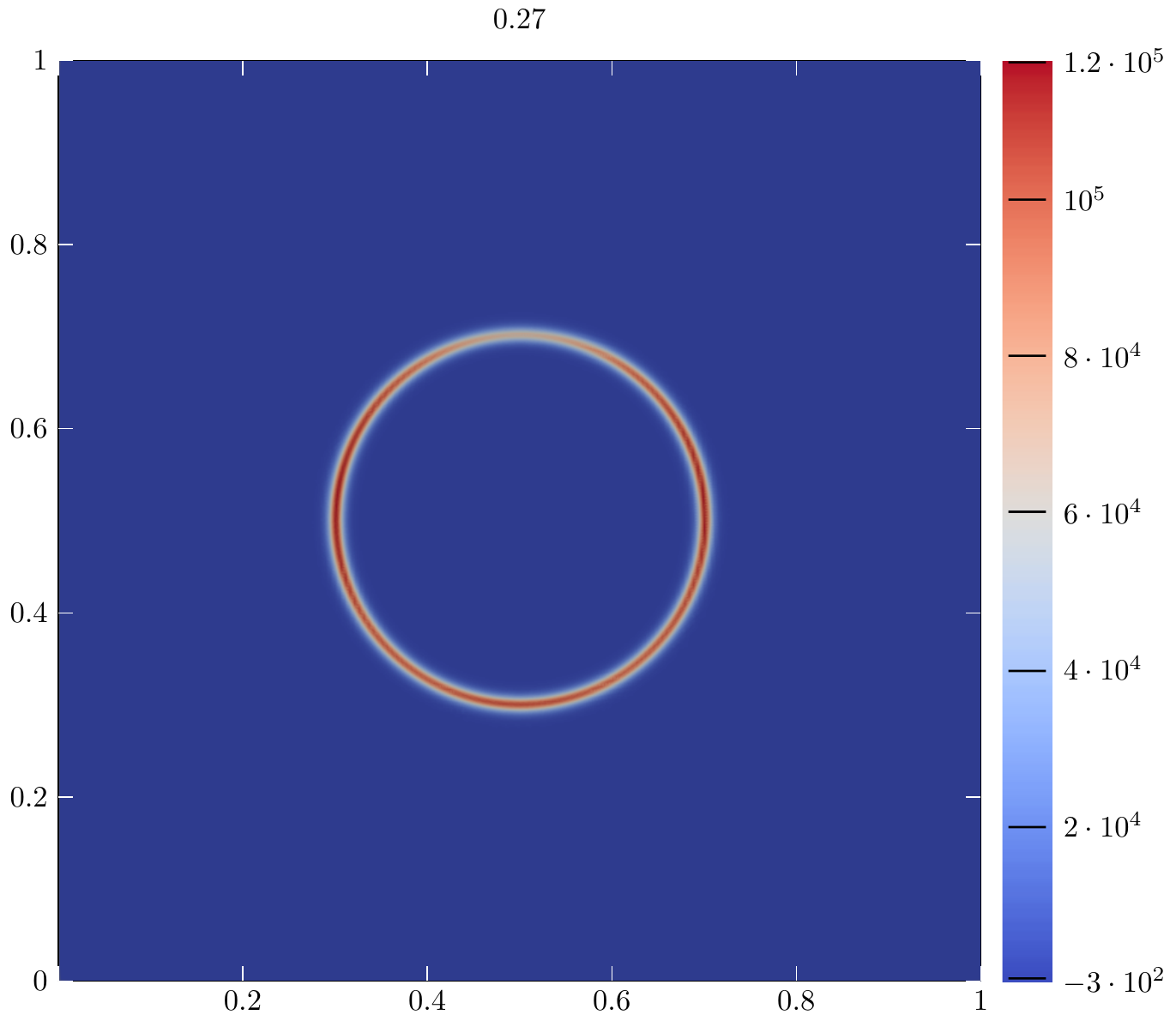}
                    \hfill
                    \includegraphics[scale=.35]{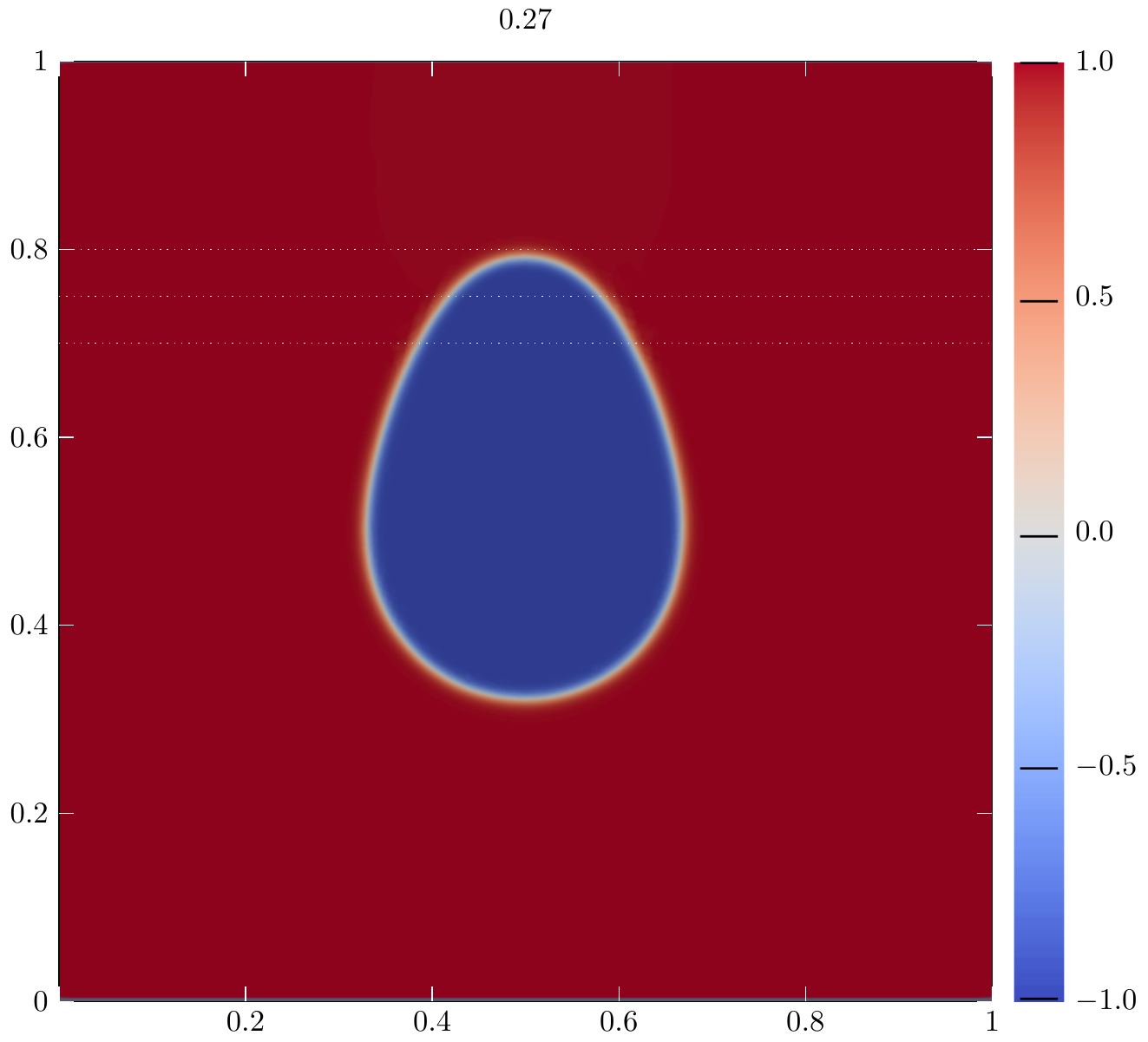}
                    \caption{Linker densities and corresponding membrane deformations
                    with $ \elComprMod = 5\cdot10^{-9}~\SIJoule\,\SIMeter^{-2}$,
                    force $ \extForceStokesSc = 500~\SIPascal\,\SIMeter^{-1} $,
                     total linker density $ \totalSpecDens = 2.5\cdot10^5~\SIMeter^{-3} $.}
                    \label{fig:linker disc high force}                    
               \end{subfigure}
                \caption{Linker disconnection during bleb expansion.}
            \end{figure}       
            Since we do not include a barrier where the 
            cortex is, the membrane is slightly pulled inwards when the bleb forms at 
            its northern front thus we have linker disconnection everywhere. 
            However, the linker disconnection at the site where the bleb develops is strongest.
            To make this effect eminently visible, we also simulated the same situation with 
            a higher force, see Figure~\ref{fig:linker disc high force}.

\begin{figure}
                \centering
                    \includegraphics[scale=.28]{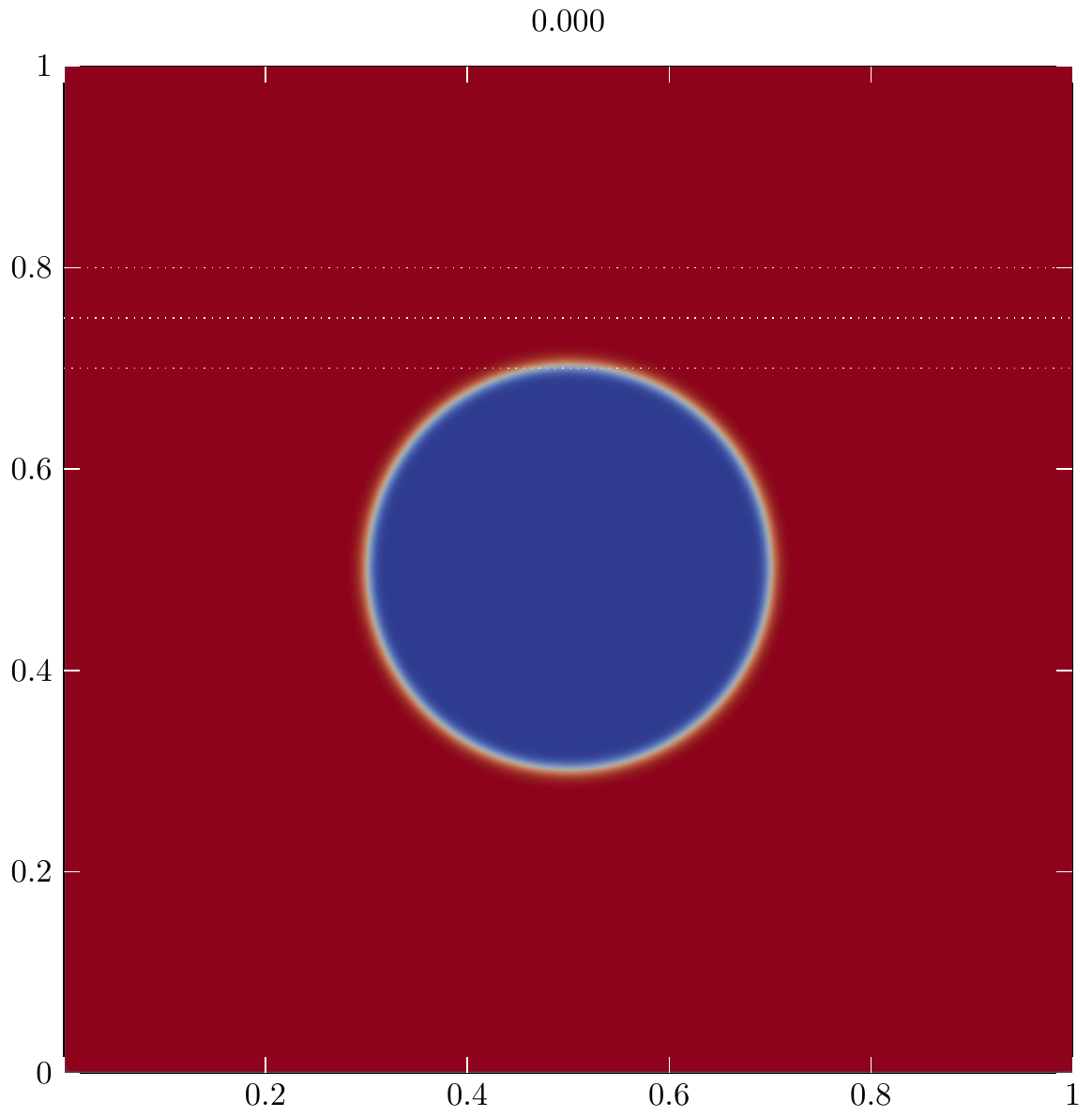}
                    \includegraphics[scale=.28]{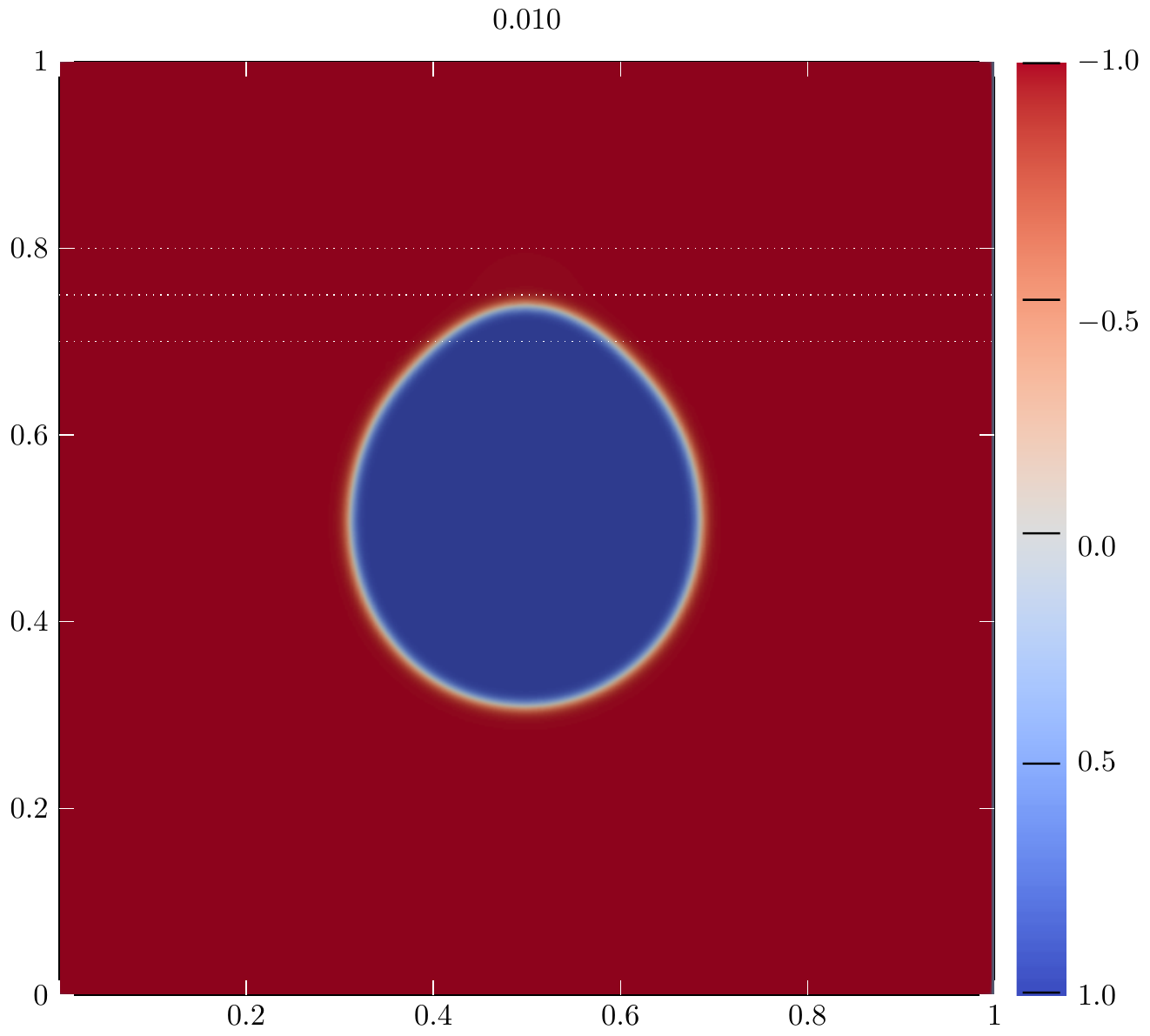}
                    \includegraphics[scale=.28]{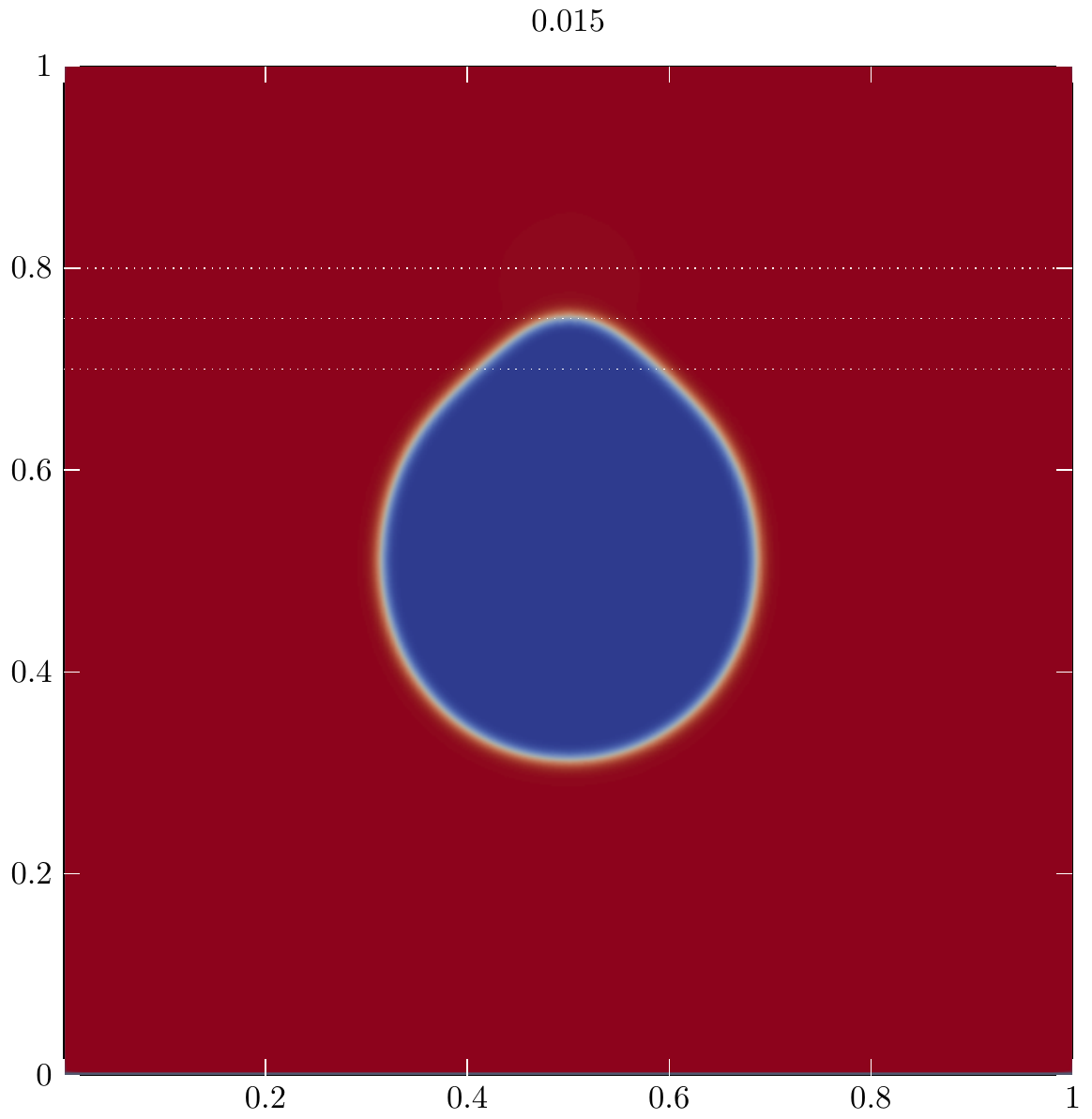}
                    \includegraphics[scale=.28]{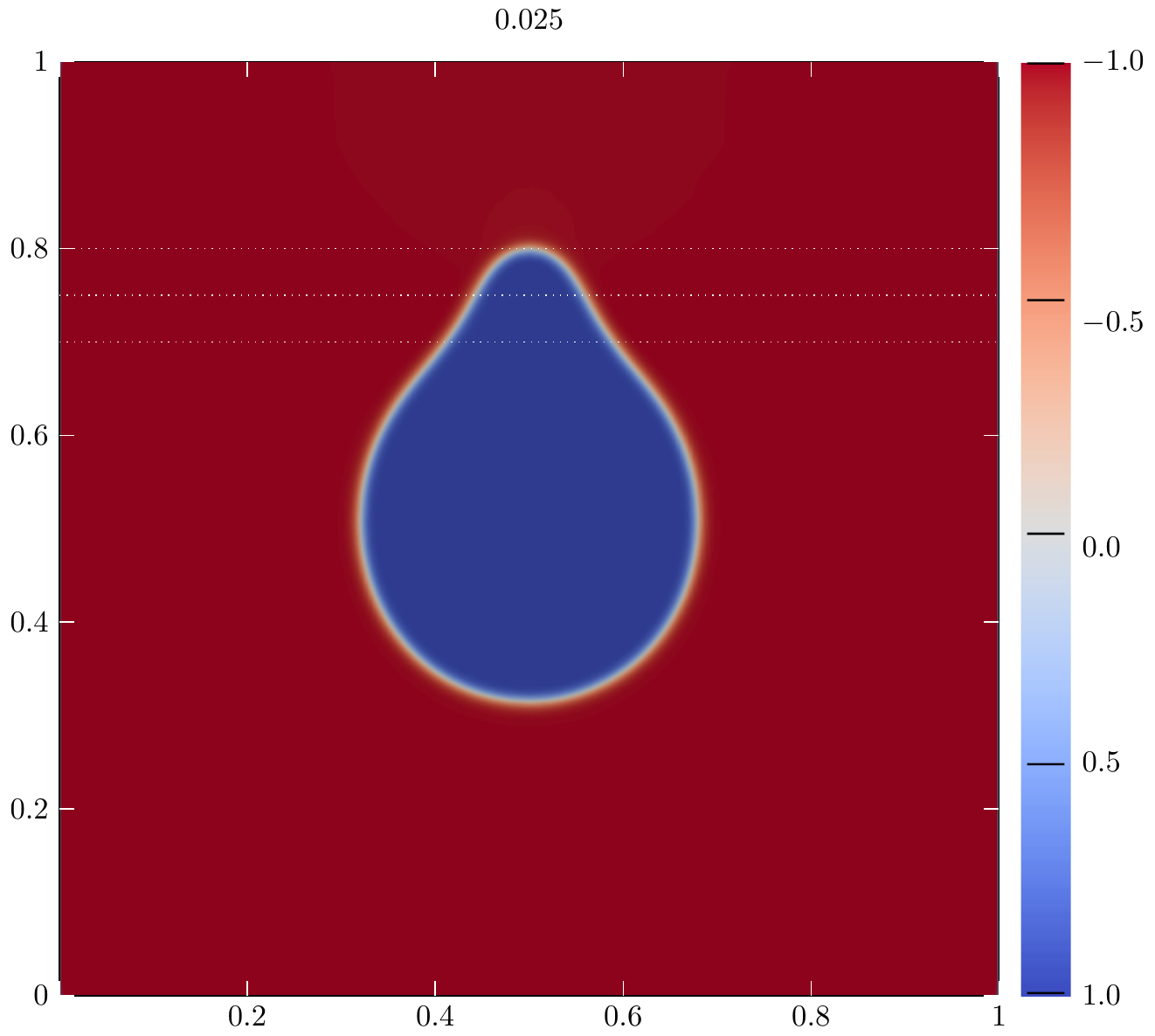}
                \caption{Bleb nucleation and expansion due to cortex destruction.}
                \label{fig:effect of the linkers forces:bleb development}
            \end{figure}

        \paragraph{Cortex Destruction}
        	Let us turn to the second scenario. The homogeneous force density 
            applied is the function in \eqref{equ:force density} without the 
            first $ \exp $ function factor. The surface tension is chosen as before.
            We consider active and inactive linkers without disconnection, but non-zero
            reconnection rate as in Table~\ref{tab:numerical methods:applications:parameter choice}.
            In the course of time, we can see a membrane protrusion developing at the site where the 
            cortex is damaged in
            Figure~\ref{fig:effect of the linkers forces:bleb development}. 
            It is interesting to note that the shape of the bleb is different from
            what we observed in 
            \autoref{fig:directed force}
            by reaching approximately the same height.
            Also, the bleb expands about $ 40 $ times faster, so for further studies one might 
            reconsider the choice of the homogeneous force's magnitude $ \extForceStokesSc $.

\section{Conclusion}
	In this work, we presented a diffuse interface model for modelling the phenomenon of ``cell 
    blebbing''. We extended the existing theory by integrating several effects considered in 
    distinct models separately into one three-dimensional model. The theoretical foundation
    of our modeling approach is Onsager's variational principle. In our free energy, we incorporated
    the well-known Canham--Helfrich energy as well as a new generalized Hookean energy that 
    accounts for the coupling of the cell membrane and the cell cortex via linker proteins that 
    can also be interpreted as surfactants. An unconditionally energy stable
    numerical scheme for space and time discretization 
    of the model with static cortex has been implemented. High computational costs 
    introduced by non-local effects from the coupling have been mitigated by a hybrid parallel 
    approach combining MPI multithreading. We then validated our modeling approach 
    numerically by qualitatively reproducing behavior of cell blebs reported in biological literature.
    
    The behavior of the protein linkers has been modeled on a pure mechanical basis in contrast
    to the dynamics of surfactants that ``actively'' participate in energy minimization. 
    Investigations in this kind of models could shed light on new aspects of linker dynamics 
    since to the best of the authors' knowledge, they have never before been considered as surfactants.
    
    To make simulation more feasible, it would be of high value to spare distributed memory 
    parallelization.
    If the connectivity of the linkers is concentrated like in our proposed Gaussian model,
    the non-local integral terms can be approximated by local terms (concentrated on the diffuse 
    layer of the cortex). An approach towards using this locality for sorting out
    large parts of the mesh efficiently during assembly might be
    a quadtree-organized mesh: modern GPUs provide dedicated 
    hardware for traversing such data structures (e.g. RT Cores or Ray Accelerators).

\section*{Acknowledgments}
The authors gratefully acknowledge the
support by the RTG 2339 ``Interfaces, Complex Structures, and Singular
Limits'' of the German Science Foundation (DFG).

\bibliographystyle{siamplain}
\bibliography{library}

\end{document}